\documentclass[letterpaper, 11pt]{article}
\usepackage{graphicx,epstopdf,subfigure,mathtools,mathrsfs, arydshln, amsmath, amssymb} 
\usepackage[font=small,labelfont=bf]{caption}
\usepackage{float}
\usepackage{authblk}
\usepackage[title]{appendix}
\PassOptionsToPackage{usenames,dvipsnames}{xcolor}
\usepackage[usenames,dvipsnames]{xcolor}
\usepackage[margin=1in]{geometry}
\usepackage[normalem]{ulem}

\usepackage{amsfonts}
\newcommand{\rone}[1]{\color{black}#1\normalcolor}
\newcommand{\roneb}[1]{\color{black}#1\normalcolor}
\newcommand{\rtwo}[1]{\color{black}#1\normalcolor}
\newcommand{\rthr}[1]{\color{black}#1\normalcolor}

\newcommand{\V}[1]{\boldsymbol{#1}}                 
\newcommand{\M}[1]{\boldsymbol{#1}}

\newcommand{\widebar}[1]{%
   \hbox{%
     \vbox{%
       \hrule height 0.5pt 
       \kern0.5ex
       \hbox{%
         \kern-0.1em
         \ensuremath{#1}%
         \kern-0.1em
       }%
     }%
   }%
}

\global\long\def\en{\V{e}_n}
\global\long\def\eb{\V{e}_b}
\global\long\def\norm#1{\left\Vert #1\right\Vert }
\global\long\def\tt#1{#1_\text{f}}
\global\long\def\tr#1{#1_\text{n}}
\global\long\def\rt#1{#1_\text{f}}
\global\long\def\rr#1{#1_\text{n}}
\global\long\def\Xs{\V{\tau}}

\global\long\def\Slet#1{\mathbb{S}\left(#1\right)}
\global\long\def\Dlet#1{\mathbb{D}\left(#1\right)}

\global\long\def\EPMI{\frac{1}{8\pi\mu}}

\global\long\def\grad{\nabla}

\global\long\def\Xhat{\widehat{\V{X}}}
\global\long\def\Uhat{\widehat{\V{U}}}
\global\long\def\fhat{\widehat{\V{f}}}
\global\long\def\rhat{\widehat{\V{r}}}

\global\long\def\ds#1{\partial_s #1}

\global\long\def\eps{{\hat{a}}}
\global\long\def\epsRS{\hat{\epsilon}}
\global\long\def\rc{a}
\global\long\def\epsc{\epsilon}
\global\long\def\er{\hat{\V{e}}_\rho}
\global\long\def\uFP{\tt{\V{U}}^{\text{(FP)}}+\tr{\V{U}}^{\text{(FP)}}}
\global\long\def\Usb{\V{U}_\text{SB}}
\global\long\def\Omsb{\Psi^\parallel_\text{SB}}

\global\long\def\Mbtt{\mathbb{M}_\text{tt}}
\global\long\def\Mbtr{\mathbb{M}_\text{tr}}
\global\long\def\Mbrt{\mathbb{M}_\text{rt}}
\global\long\def\Mbrr{\mathbb{M}_\text{rr}}

\graphicspath{{../HelicalSwimming/}}

\usepackage{hyperref}
\hypersetup{
    colorlinks=false,
    pdfborder={0 0 0},
}

\title{Slender body theories for rotating filaments\vspace{-0.5 cm}}
\author{Ondrej Maxian and Aleksandar Donev \vspace{-0.5 cm}}

\begin{document}
\maketitle

\begin{abstract}
Slender fibers are ubiquitous in biology, physics, and engineering, with prominent examples including bacterial flagella and cytoskeletal fibers. In this setting, slender body theories (SBTs), which give the resistance on the fiber asymptotically in its slenderness $\epsc$, are useful tools for both analysis and computations. However, a difficulty arises when accounting for twist and cross-sectional rotation: because the angular velocity of a filament can vary depending on the order of magnitude of the applied torque, asymptotic theories must give accurate results for rotational dynamics over a range of angular velocities. In this paper, we first survey the challenges in applying existing SBTs, which are based on either singularity or full boundary integral representations, to rotating filaments, showing in particular that they fail to consistently treat rotation-translation coupling in curved filaments. \rone{We then provide an alternative approach which approximates the three-dimensional dynamics via a one-dimensional line integral of Rotne-Prager-Yamakawa regularized singularities. While unable to accurately resolve the flow field near the filament,} this approach gives a grand mobility with symmetric rotation-translation and translation-rotation coupling, making it applicable to a broad range of angular velocities. To restore fidelity to the three-dimensional filament geometry, we use our regularized singularity model to inform a simple empirical equation which relates the mean force and torque along the filament centerline to the translational and rotational velocity of the cross section. The single unknown coefficient in the model is estimated numerically from three-dimensional boundary integral calculations on a rotating, curved filament.
\end{abstract}

\section{Introduction}
Whether an organism has one or millions of cells, slender filaments play a vital role in its motion and maintenance. For instance, these types of filaments make up the flagellar appendages that bacteria use to propel themselves in run and tumble motions \cite{berg2004coli, lauga09rev}. In animal cells, slender filaments are a key component in the cell cytoskeleton, which provides a scaffolding for the cell \cite{alberts}, and adapts to promote cellular division, migration, and resistance to deformation \cite{eghiaian2015}. 

Given the vital role of slender filaments in biology, it is not surprising that their interaction with their surrounding fluid medium has been a subject of study in physics and applied mathematics. Because most of the applications of interest involve small lengthscales, the relevant fluid equations are the Stokes equations, which are linear and therefore possible to solve analytically through the use of the Stokes Green's function. However, because the body geometry is slender, it is expensive and often intractable to solve the Stokes equations on the surface of the slender filament. This intractability led Batchelor to develop the first so-called ``slender body theory'' (SBT) \cite{batch70}, in which the body is approximated via a one-dimensional line of Stokeslet singularities. Subsequent work by Keller and Rubinow \cite{krub} and Johnson \cite{johnson} placed Batchelor's work on more rigorous asymptotic footing; in particular, Johnson \cite{johnson, johnsonTh} showed that adding a doublet to the line integral produces a velocity that is guaranteed to be constant to $\mathcal{O}(\epsc)$ on the fiber cross section, where $\epsc$ is the filament slenderness.

Recent work has focused on placing these theories back in the context of three-dimensional well-posed partial differential equations (PDEs) \cite{morifree, mori2020accuracy, mori2018theoretical, koens2018boundary}. In particular, it has been shown that SBT can be derived from a three-dimensional boundary integral equation \cite{koens2018boundary, gargslender}, and that its solution is an $\mathcal{O}(\epsc)$ approximation to a Stokes PDE with mixed Dirichlet-Neumann boundary data and a boundary condition that the fiber maintain the integrity of its cross section \cite{morifree, mori2020accuracy, mori2018theoretical}. A second class of recent work has developed SBTs for different geometries, including ribbons \cite{koens2016slender} and bodies with non-circular cross section \cite{borker2019slender}. 

Despite the breadth of literature on SBT and the novel theories for complex geometries, there remains an open problem which is simple to formulate, but difficult to solve, having to do with the \emph{rotation} of a filament. Consider a filament with one clamped end and one free end, where a motor spins the clamped end with frequency $\omega$. At some critical frequency $\omega_c$, the applied torque overcomes the bending moment, the steady twirling state becomes unstable, and large-scale filament deformations are introduced. This instability, which has been studied extensively in \cite{wolgemuth2000twirling,lim2004simulations,wada2006non,lee2014nonlinear,
maxian2022hydrodynamics} has applications in the motion of bacterial flagella \cite{chen2000torque, yuan2010asymmetry} and supercoiling during DNA transcription \cite{liu1987supercoiling}. It has been shown that the critical frquency $\omega_c$ scales like $1/\epsc^2$, and thus that the torque required to induce it is approximately independent of the fiber aspect ratio \cite[Eq.~(62)]{powers2010dynamics}. Because of this, fibers being driven at the critical frequency have angular velocities which are sufficiently large as to also affect the translational dynamics. But as of yet, we are not aware of any slender body theory that can consistently account for the full hydrodynamics of a slender, translating, \emph{and} rotating filament, and in particular for the coupling between the two. In fact, we recently showed that the failure to account for rotation-translation coupling causes the SBT-based analysis of \cite{wolgemuth2000twirling} to overestimate $\omega_c$ by as much as 20\% \cite[Sec.~5.3]{maxian2022hydrodynamics}. 

To understand the reason for the failures of previous SBTs, in Section\ \ref{sec:SBT3D} we present a survey of an old and new approach to SBT, and an explanation of the challenges involved when we try to extend these approaches to account for rotation. \rone{Fundamentally, the issues that we point out can be traced to the difference between the \emph{resistance} problem, which seeks the force leading to a given velocity, and the \emph{mobility} problem, which is the inverse of the resistance problem and has different asymptotic behavior. The problem that we consider here is a \emph{mobility} problem, since the torque driving the filament is constant while the angular velocity changes with $\epsc$. It is therefore not necessarily surprising that prior SBTs, which were designed with the resistance problem in mind, cannot be applied to the spinning filament mobility problem. 

One potential workaround to this is the recent approach of Koens \cite{koens2022tubular}, which is designed to \emph{exactly}  solve the resistance (and therefore mobility) \roneb{problem by solving a series of Fredholm integral equations of the second kind to find the surface traction. These integral equations are similar to that of SBT, but are instead forced by a function of the previous order traction. The approach, however, offers little insight into the local resistance behavior under high rotation, and it is unclear how many iterations would be required to accurately capture the nonlocal translation-rotation coupling that is not correctly treated by SBT.} For this reason, we seek an alternative approximation which takes advantage of the fiber slenderness to minimize the number of nonlocal integral evaluations and allow for dynamic simulations with many fibers.}

In Section\ \ref{sec:RPYSBT} of this paper, we offer a line integral of regularized singularities as a potential alternative to traditional SBT. Mathematically, the approach of the original slender body pioneers \cite{batch70, krub, johnson} was to distribute singularities along the centerline, then evaluate the flow from these singularities on a fiber cross section. The flow is nonsingular since it is \emph{evaluated} a fiber radius away from the location of the singularities. In regularized singularities, we instead distribute the force and torque over a ``cloud'' or ``blob'' region of characteristic size $\eps \sim \rc$, which is defined by some regularized delta function $\delta_\eps$. \rthr{Throughout this paper, $\delta_\eps$ will be a surface $\delta$ function on a sphere of radius $\eps$, but other approaches are also possible}. In this case, the velocity field is nonsingular because the force and torque are \emph{distributed} over a region of finite size approximately equal to the fiber radius. We can then convolve the velocity field (or half the vorticity field) with the regularized delta function again to obtain the velocity (or angular velocity) of the fiber centerline. In this way, we derive fundamental regularized singularities for all components of the mobility operator, then perform asymptotics on line integrals of these regularized singularities. \rone{The sum of these components gives a \emph{model} of the filament dynamics with the correct physical behavior that can be used in practice without approximating precisely a specific geometry; after all, the geometry of an actin filament \cite{alberts} or flagella \cite[Fig.~5.3]{berg2004coli} is not perfectly spheroidal or even tubular, and so one could argue that the regularized singularity model is just as much of an approximation as SBT.}

Representing filaments using regularized singularities is not a novel idea, as it was first employed for numerical purposes in the immersed boundary (IB) method of the early 1970s to model the flow patters inside the heart \cite{peskin1972flow}. Later variations, including the force coupling method \cite{fcm03}, method of regularized Stokeslets \cite{cortez2001method, cortez2005method}, and Rotne-Prager-Yamakawa (RPY) tensor \cite{RPYOG, wajnryb2013generalization}, focused on choosing the regularized delta function such that the Stokes equations are solvable analytically. \rtwo{While not a unique or necessarily superior choice, we prefer to use the RPY kernel because of its simple far-field analytical form (as a Stokeslet plus a source doublet), its symmetric positive definite property, and because of its long history of use in polymer suspensions (e.g., \cite{butler2005brownian, beck2006ergodicity, wang2012flipping, keavRPY}). The analysis here could be conducted for any regularization function where the mobility between two points can be determined analytically.}

Recently, the accuracy of using regularized singularities to approximate slender \emph{translating} filaments has come under scrutiny both from a numerical \cite{ttbring08} and analytical \cite{cortez2012slender, maxian2021integral, ohm2021remarks, zhao2021regularized} perspective. \rone{Generally speaking, the recent analysis has affirmed \cite{cortez2012slender, maxian2021integral, ohm2021remarks} that regularization methods can effectively match the slender body theory \emph{centerline and far-field fluid velocity} with a judicious choice of regularization radius $\eps$, while others \cite{ ohm2021remarks} have shown that the global flow field \emph{near} the slender body cannot be matched unless very specific conditions on the regularization function are met \cite[Sec.~4]{zhao2021regularized}.} \rtwo{While the latter restriction is somewhat disappointing, it has little effect in practice: since regularized singularities can match the SBT filament centerline velocity, the only impact of the flow near the filament being incorrect is in computing disturbance flows on other nearly touching filaments. These disturbance flows change sharply over distances of $\mathcal{O}(\epsilon)$, which makes a one-dimensional SBT approach for nearly touching fibers quite challenging if not impractical.} One of the goals of this paper is therefore to determine how the \emph{centerline} motion induced by \rthr{RPY} singularities compares to that of SBT when rotation is considered. For instance, do they reproduce the relationship between torque and rotational velocity for a straight cylinder? And what about the feedback of rotational velocity on translational motion for curved filaments? 

These are the main questions we try to answer in this paper. To do so, we first look at how far we can get with traditional SBT in Section\ \ref{sec:SBT3D}. We show that the singularity methods of Keller and Rubinow and Johnson yield well-defined mobilities for translation from force (trans-trans) and rotation from torque (rot-rot) terms, but that both leave an unknown $\mathcal{O}(1)$ constant in the rotation-translation (rot-trans) coupling term which is relevant \rone{in the mobility problem}. Then, we perform asymptotic expansions on line integrals of regularized RPY singularities to derive a consistent SBT for filaments rotating and translating with speeds of arbitrary order. We show that the regularized singularity radius $\eps$ can be chosen to match either the \rthr{three-dimensional} trans-trans \emph{or} rot-rot mobility, but not both, \rthr{which implies that the set of RPY singularities we consider cannot accurately model a three-dimensional cylinder}. We also show that, unlike SBTs that are more faithful to the filament geometry, the regularized RPY singularity approach gives symmetric rot-trans coupling terms, and naturally extends to the fiber endpoints.

The main drawback of our regularized singularity approach is that we lose fidelity to the true tubular geometry of the filament. However, since we expect the regularized singularity approach to capture the essential physics of the interaction between the filament and fluid, we can utilize it as a starting point to model the three-dimensional dynamics. To this end, in Section\ \ref{sec:BI} we use our results from Section\ \ref{sec:RPYSBT} to postulate an asymptotic relationship between the force and torque on a filament's cross section and its translational and rotational velocity. This equation has a single unknown constant, which we estimate through three-dimensional boundary integral simulations, thereby developing an empirical SBT for tubular rotating filaments.

\section{Three-dimensional SBTs for twist \label{sec:SBT3D}}
In this section, we discuss how previous SBTs for translation might be extended to also account for rotation. This section is not meant to be a comprehensive list of previous SBTs (see \cite{SBTRev} for a such a review), but rather a focus on two different techniques and the challenges involved in applying them to a (rapidly) \emph{rotating} filament. Our first technique is that used by Keller and Rubinow \cite{krub} and Johnson \cite{johnson, johnsonTh}, which is to represent the flow field by an integral of singularities positioned on the filament centerline. The approach of Keller and Rubinow is technically distinct from that of Johnson since they only use the singularity representation in the outer expansion, but the end result is the same, and so we discuss both in Section\ \ref{sec:johnson}. We then discuss the more recent approach of Koens and Lauga \cite{koens2018boundary}, which evaluates the three-dimensional boundary integral representation of the surface velocity asymptotically in $\epsilon$. The asymptotics of this approach are more involved, since it maps traction on a surface to velocity on a surface, rather than singularities on a line to velocity on the surface. 

To establish some notation, we let $\V{X}(s)$ be the fiber centerline, where $s \in [0,L]$ is an arclength parameterization of the curve, and thus $\Xs(s)=\ds{\V{X}}(s)$ is the unit-length tangent vector. The fiber surface $\widehat{\V{X}}$ is then defined as
\begin{equation}
\label{eq:Xhat}
\widehat{\V{X}}(s,\theta)=\V{X}(s)+\rc \rho(s) \left(\en(s) \cos{\theta}+\eb(s) \sin{\theta}\right):=\V{X}(s)+a \rho(s) \widehat{\V{r}}(s,\theta),
\end{equation}
where $a\rho(s)$, $0 \leq \rho(s) \leq 1$, is the radius of the (circular) cross section at $s$, $\en(s)=\V{X}_{ss}(s)/\kappa(s)$ is the fiber normal, $\kappa(s)$ is the centerline curvature at $s$, and $\eb = \Xs \times \en$ is the binormal. The maximum fiber radius is $\rc$, which gives an aspect ratio $\epsc:=\rc/L$. Unless otherwise specified, we will assume that we are working in a region of the fiber where $\rho(s) \equiv 1$; thus $\rc \rho=\rc$. 

We will assume that the fiber centerline is translating with velocity $\V{U}(s)$ while its cross section rigidly rotates with angular velocity $\Psi^\parallel(s)$ around the tangent vector $\Xs(s)$, so that the velocity on the surface of the fiber is given by
\begin{equation}
\label{eq:Uwant}
\Uhat(s,\theta) = \V{U}(s) + \rc \Psi^\parallel(s) \left(\Xs(s) \times \widehat{\V{r}}(s,\theta)\right) 
\end{equation}
 The goal of our analysis is to find an equation relating this motion to the total force density $\V{f}(s)$ and parallel torque density $n^\parallel(s)$ on the fiber cross section. We assume here that the no-slip condition applies to the fluid velocity $\V{u}(\V{x})$, so that 
\begin{equation}
\label{eq:UMatch}
\V{u}\left(\Xhat(s,\theta)\right)=\Uhat(s,\theta).
\end{equation}

\subsection{Singularity representations \label{sec:johnson}}
We first try to apply an older SBT approach to the spinning filament problem: the singularity approach of Johnson \cite{johnson}, and later G\"otz \cite{gotz2001interactions}. The approach of Johnson is to look for a series of singularities along the fiber centerline to represent the flow field globally (everywhere outside the fiber). The singularities are chosen so that the boundary condition\ \eqref{eq:UMatch} is satisfied with an error $\mathcal{O}(\epsc)$. 

Using superposition, we can break the boundary velocity\ \eqref{eq:Uwant} into a velocity with $\V{U}=\V 0$ and another with $\Psi^\parallel=  0$. For $\Psi^\parallel= 0$, Johnson shows that the boundary condition\ \eqref{eq:UMatch} can be satisfied to $\mathcal{O}(\epsc)$ by setting the global flow outside of the fiber to be an integral of Stokeslets and doublets,
\begin{gather}
\label{eq:u0}
\V{u}^{(u)}(\V{x}) = \EPMI \int_0^L \left(\mathbb{S}\left(\V{x},\V{X}(s)\right)+\frac{\left(a \rho(s)\right)^2}{2}\mathbb{D}\left(\V{x},\V{X}(s)\right)\right)\V{f}(s)  \, ds \\ \nonumber
\text{where} \qquad \mathbb{S}\left(\V{x},\V{X}(s)\right) = \left(\frac{\M{I}}{r}+\frac{\V{r}\V{r}^T}{r^3}\right),\qquad \qquad \mathbb{D}\left(\V{x},\V{X}(s)\right) = \left(\frac{\M{I}}{r^3}-3\frac{\V{r}\V{r}^T}{r^5}\right),
\end{gather}
\rone{$\mu$ is the background fluid viscosity}, and $\V{r}=\V{x}-\V{X}(s)$ with $r=\norm{\V{r}}$. The doublet strength is necessary to cancel the $\mathcal{O}(1)$ angular-dependent flow induced by the Stokeslet on a filament cross section. Since the doublet has strength $\mathcal{O}(\epsc^2)$, it makes a negligible contribution to the outer expansion, but corrects the Stokeslet flow in the inner expansion. The resulting asymptotic evaluation of $\V{u}^{(u)}$ on the filament cross section $\widehat{\V{X}}(s,\theta)$ is independent of $\theta$ to $\mathcal{O}(\epsc)$, and is given by
\begin{gather}
\label{eq:uUexp}
8 \pi \mu \V{u}^{(u)}\left(\widehat{\V{X}}(s,\theta)\right) =\tt{\V{U}}^\text{(LD)}(s)+\tt{\V{U}}^{\text{(FP)}} + \mathcal{O}(\epsc), \\[4 pt] 
\label{eq:ULD}
\tt{\V{U}}^\text{(LD)}(s)=\ln{\left(\frac{4s(L-s)}{\rc^2}\right)} \left(\M{I}+\Xs(s) \Xs(s)\right) \V{f}(s) + \left(\M{I}-3\Xs(s) \Xs(s)\right) \V{f}(s)\\[4 pt] 
\label{eq:UfFP}
\tt{\V{U}}^{\text{(FP)}}(s) = \int_0^L \left(\mathbb{S}\left(\V{X}(s),\V{X}(t)\right)\V{f}(t)-\left( \frac{\M{I}+\Xs(s) \Xs(s)}{|t-s|}\right)\V{f}(s)\right)
\end{gather}
where the doublet contributes the $\left(\M{I}-3\Xs \Xs\right)$ term \cite[Eqs.~(10--12)]{johnson}, and the term $\tt{\V{U}}^{\text{(FP)}}$ is a finite part integral which is only well-defined as a difference of two terms (the outer and inner expansions). The expression\ \eqref{eq:uUexp} has an error on the cross section of order $\epsc$, which means there will be a nonzero $\mathcal{O}(1)$ rotational velocity of the cross section. To zero out the rotational velocity, Johnson shows that it is necessary to introduce a rotlet, a source, two stresslets, and two quadrupoles, with strengths given as a function of the $\mathcal{O}(\epsc)$ velocity generated by the surface flow\ \eqref{eq:uUexp}. While these singularities zero out the $\mathcal{O}(\epsc)$ term, they \emph{do not} impart any additional translational velocity on the cross section, meaning that\ \eqref{eq:uUexp} gives the \emph{translational} velocity on the cross section to order $\epsc^2 \log{\epsc}$.

Let us now consider how the fluid flow outside the fiber relates to its angular velocity. Johnson first sets
\begin{equation}
\label{eq:Jnpar}
n^\parallel = 4 \pi \mu \rc^2 \Psi^\parallel,
\end{equation}
which is the result for an infinite, straight cylinder \cite[Eq.~(62)]{powers2010dynamics}. The singularity representation Johnson uses to match the angular velocity everywhere is then given by a line integral of rotlets, 
\begin{gather}
8 \pi \mu \V{u}^{(\Psi)}\left(\V{x}\right)=\int_0^L  \mathbb{R}\left(\V{x},\V{X}(s)\right)n^\parallel(s) \, ds, \qquad 
 \mathbb{R}\left(\V{x},\V{X}(s)\right)=  \frac{\Xs(s) \times \V{r}}{r^3}
\end{gather}
Asymptotically, this velocity is equal to 
\begin{gather}
\label{eq:uOmEx}
8\pi \mu \V{u}^{(\Psi)}\left(\Xhat\left(s,\theta\right)\right) \approx 
\frac{2 n^\parallel}{\rc}\left(\Xs \times \hat{\V{r}}\right) +\frac{n^\parallel}{2} \left(\ln{\left(\frac{4s(L-s)}{ \rc^2}\right)}-2\right) \left(\Xs \times \ds \Xs\right)+\tr{\V{U}}^{\text{(FP)}}\\ \nonumber 
 +n^\parallel \kappa \cos{\theta}\left(\Xs \times \hat{\V{r}}\right) +
\mathcal{O}\left(\epsilon \log{\epsilon}\right), \\ 
\label{eq:UnFP}
\text{where} \qquad \tr{\V{U}}^{\text{(FP)}}(s)=\int_0^L \left(\mathbb{R}\left(\V{X}(s),\V{X}(t)\right)n^\parallel(t) - \frac{1}{2}\left(\frac{\Xs(s)\times\ds{\Xs}(s) }{|t-s|}\right)n^\parallel(s)\right)
\end{gather}
on the fiber surface. Substituting\ \eqref{eq:Jnpar} into\ \eqref{eq:uOmEx}, we see that the first term is exactly the angular velocity of\ \eqref{eq:Uwant}, while the second and third terms on the first line of\ \eqref{eq:uOmEx} give an additional constant velocity on the cross section (the rot-trans coupling term). However, we still have an angular-dependent term on the second line of\ \eqref{eq:uOmEx}: since $\Xs \times \hat{\V{r}}=\eb \cos{\theta}  - \en \sin{\theta} $, the term in the second line has the angle dependence 
\begin{align}
n^\parallel \kappa \cos{\theta}\left(\Xs \times \hat{\V{r}}\right)&=
\label{eq:ExpandAng}
 \frac{n^\parallel \kappa}{2} \left( \eb +\eb \cos{2\theta} -\en \sin{2\theta}\right).
\end{align}

 \rone{In Johnson's analysis, which is based on the resistance problem, the angular velocity $\Psi^\parallel$ and translational velocity $\V{U}$ are independent of $\epsc$, and so the contribution of all subleading terms in\ \eqref{eq:uOmEx} to translational velocity on the cross section is still $\epsc^2$ relative to the contribution of forcing.} However, if force and torque are of the same order in $\epsc$ (as \rone{in the mobility problem} when a fiber is driven by a torque $n^\parallel$ independent of $\epsc$), the second line will make an $\mathcal{O}(1)$ contribution to the cross-sectional velocity, violating the boundary condition\ \eqref{eq:UMatch}. 
Thus, there is a $\mathcal{O}(1)$ angle dependence coming from the rotlet term which is unresolved. This dependence is the reason why the final integral equation of Keller and Rubinow \cite[Eq.~(28)]{krub} for twisting filaments contains terms which depend on the angle $\theta$ at which the inner expansion (of a rotating, translating cylinder) is matched with the outer expansion (of Stokeslets and rotlets), meaning there is no general solution for the Stokeslet and rotlet strength which is independent of the matching angle. 

In addition, there is also a lack of symmetry in the final result of Johnson, since torque makes an $\mathcal{O}(\log{\epsc})$ contribution to $\V{U}$ in\ \eqref{eq:uOmEx}, but force does not contribute to $\Psi^\parallel$ in\ \eqref{eq:Jnpar}. In our prototypical example of an unstable twirling fiber, this is not an issue, since force makes a negligible contribution to rotational velocity anyway.\footnote{Specifically, if the motor is driven by a constant torque $n^\parallel$, the angular velocity $\Psi^\parallel \sim n^\parallel/\epsc^2$. Symmetry tells us that the contribution of force to $\Psi^\parallel$ is $\mathcal{O}(\log{\epsc})$, which means that the rotational velocity from force is $\rt{\Psi}^\parallel  \sim f \log{\epsc}$, which is much smaller than the rotational velocity from torque $\rr{\Psi}^\parallel \sim n/\epsc^2$, if $f$ and $n$ scale similarly with $\epsc$.} In the case when $\Psi^\parallel$ does not scale with $\epsc$, however, the torque $n^\parallel$ is order $\epsc^2$ and therefore makes a negligible contribution to $\V{U}$, but the force might make a nontrivial contribution to $\Psi^\parallel$. Thus there are two main issues when applying singularity representations to twisting filaments: first, the representation of the rotation in terms of rotlets is insufficient to satisfy the boundary condition\ \eqref{eq:UMatch} to $\mathcal{O}(\epsc)$, and, second, there is a lack of symmetry in the rot-trans coupling terms.

\subsection{Boundary integral representations \label{sec:KL}}
Recently, effort has been made to place the asymptotic theories of Keller and Rubinow and Johnson on a more rigorous three-dimensional footing. Foremost among these are Mori et al. \cite{mori2018theoretical, morifree, mori2020accuracy}, who show that slender body theory is an $\mathcal{O}(\epsc)$ approximation of a three-dimensional PDE with non-standard boundary conditions, and Koens and Lauga \cite{koens2018boundary}, who show that the translational SBT of Johnson and Keller and Rubinow can be obtained from asymptotic expansion of the single layer\footnote{Koens and Lauga begin with the full boundary integral representation, but the single layer representation is sufficient as long as there is no change in the fiber volume. However, it is important to point out that $\fhat$ in\ \eqref{eq:UBI} is not in general the force density on the fiber surface. Notable exceptions are a filament undergoing rigid body motion or a ``filament'' whose interior is a fluid of viscosity equal to that of the fiber exterior.} boundary integral equation \cite[Sec.~4.1]{pozrikidis1992boundary}
\begin{equation}
\label{eq:UBI}
8 \pi \mu \, \Uhat\left(s,\theta\right) = \int_0^{2\pi} \int_0^L \Slet{\Xhat(s,\theta),\Xhat (s',\theta')} \fhat\left(s',\theta'\right)\, ds' d\theta'.
\end{equation}
Here we assume the fiber is a thin shell filled with fluid, so that $\fhat(s,\theta)$, which is the jump in surface traction times the surface area element \cite[p.~105]{pozrikidis1992boundary}, is also equal to the force density on the fiber surface. Koens and Lauga show that the constant cross-sectional $\V{U}(s)$ in\ \eqref{eq:Uwant} is related to the total force on a cross section,
\begin{equation}
\label{eq:fTrac}
\V{f}(s) = \int_0^{2 \pi}\fhat\left(s,\theta\right) \, d\theta
\end{equation}
by the classical SBT\ \eqref{eq:uUexp}, and that, for a \emph{straight} cylinder, the torque-angular velocity relationship\ \eqref{eq:Jnpar} is recovered \cite[Eq.~(4.11)]{koens2018boundary} for the total parallel torque on a cross section,
\begin{equation}
\label{eq:nparTrac}
n^\parallel(s) =  \Xs(s) \cdot \int_0^{2\pi} a\rho(s) \left(\rhat(s,\theta) \times \fhat(s,\theta)\right) \, d\theta.
\end{equation}

Here we briefly discuss how the approach of Koens and Lauga might be extended to account for rotation in curved filaments. The main issue is that, if the boundary integral equation\ \eqref{eq:UBI} is only expanded to leading order (i.e., with an error of $\mathcal{O}(\epsc)$), it is impossible to make observations about the rotational velocity, since the velocity on the cross section induced by an angular velocity $\Psi^\parallel$ is $\sim \epsc \Psi^\parallel$. Thus more terms are necessary in the asymptotics. Unfortunately, because the expansions must be conducted around both $\theta$ and $s$ in\ \eqref{eq:UBI}, even expanding a single layer formulation to the next order is complex, \rone{as can be seen in the results of slender phoretic theory \cite{katsamba2020slender}, which completely extends the approach of Koens and Lauga to next order for the case of a body which swims via a fluid slip along a self-generated surface chemical concentration gradient. }

To get a flavor for the challenges involved, we consider an asymptotic expansion of the \emph{isotropic part} of the single-layer potential (c.f. \cite[Eq.~(5.1)]{koens2018boundary}
\begin{equation}
\label{eq:Ui1B4}
\V U_{i1}(s,\theta):=\int_{0}^{2\pi}  \int_{0}^L \frac{\V{f}(s',\theta')}{R\left(\widehat{\V{X}}(s,\theta),\widehat{\V{X}}(s',\theta')\right)} \, d\theta' \, ds',
\end{equation}
in an inner region where $s-s' = \mathcal{O}(\rc)$. The details of the expansion up to integration over $\theta$ are given in Appendix\ \ref{sec:KLAppen}. In order to integrate over $\theta'$, we need to assume a Fourier-series representation of $\V{f}$. For simplicity, we will look only at the first two terms, 
\begin{equation}
\label{eq:OneTrac}
\fhat(s,\theta) = \frac{1}{2\pi}\V{f}(s) + \frac{1}{2\pi}\left(\V{f}_{c,1}(s) \cos{\theta } + \V{f}_{s,1}(s) \sin{\theta }\right),
\end{equation}
which we substitute into\ \eqref{eq:Uinttheta}. After integrating over $\theta'$, the final inner expansion for the surface velocity is\footnote{We use the variable $\theta$ here for notational convenience; more precisely, $\theta$ should be measured relative to the curve torsion angle $\theta_i$(s), which is defined in Appendix\ \ref{sec:KLAppen}. That is, $\theta$ in\ \eqref{eq:OneTrac} and\ \eqref{eq:Ui1} stands for $\theta-\theta_i(s)$.}
\begin{gather}
\label{eq:Ui1}
\V U_{i1}(s,\theta) =\ln{\left(\frac{4s(L-s)}{\rc^2}\right)}\V{f}(s)+(L-2s)\ds \V{f}(s)+\V{f}_{c,1}(s) \cos{\theta}\\ \nonumber
-\frac{\rc \kappa}{2} \left(\V{f}(s)\cos{\theta} + \left(\frac{1}{4}\cos{2\theta} + \frac{1}{2}\left(\ln{\left(\frac{4s(L-s)}{\rc^2 }\right)}-2\right)\right)\V{f}_{c,1}(s)
-\frac{1}{2}\V{f}_{s,1}(s)\cos{\theta}\sin{\theta}\right)+\mathcal{O}(\epsc^2).
\end{gather}
We can make three important observations from\ \eqref{eq:Ui1}: first, the term $\V{f}(s) \kappa \cos{\theta}$ demonstrates rot-trans coupling (rotational motion from constant forcing) which results from including additional terms in the SBT expansion, or more specifically from accounting for the curvature of the fiber in the inner expansion. Likewise, the \emph{torque}\ \eqref{eq:nparTrac} generated from the traction scales as $\rc \V{f}_{c/s,1}$, and thus\ \eqref{eq:Ui1} tells us that torque makes an $\mathcal{O}(\log{\epsc})$ contribution to $\V U(s)$, as we expect from the result\ \eqref{eq:uOmEx} of singularity methods. 

While the result of Koens and Lauga is vital to understanding translational SBT, there are clearly limitations to extending their approach to rotation. Even for the simplest possible term in the boundary integral formulation (isotropic part of the single layer), there are many terms that have to be tracked if we expand to $\mathcal{O}(\epsc)$, i.e., to an error of $\mathcal{O}\left(\epsc^2\right)$. More importantly, it is not even clear if we can solve\ \eqref{eq:Ui1}, since the Fourier modes no longer decouple as they do for translation. 

Summing up this section, we have seen that a slender body theory which properly accounts for twist dynamics \emph{and} is faithful to the three-dimensional fiber geometry remains elusive. In order to make some progress, we make an approximation of the fiber geometry and define the fiber as a series of infinitely many regularized singularities. Then, we use our regularized singularity SBT to inform an empirical SBT for the full three-dimensional geometry.

\section{Slender body theory from regularized singularities \label{sec:RPYSBT}}
Our survey in Section\ \ref{sec:SBT3D} reveals the challenging nature of developing a slender body theory for twist. Because of the separation of scales between the angular and translational velocity, more terms are required in the asymptotic analysis, which makes these SBTs tedious to derive and cumbersome to work with in practice.

An alternative approach, which we consider in this section, is to relax the fidelity to a true cylindrical geometry and instead represent the cylinder using \emph{regularized} singularities distributed along the centerline. The idea of the regularized singularity approach is to replace the delta function in the force and torque with a regularized delta function $\delta_\eps$, where $\eps$ is the ``regularization radius'' that is chosen to model the three-dimensional problem. \rthr{Our choice of kernel is the Rotne-Prager-Yamakawa (RPY) tensor \cite{RPYOG, wajnryb2013generalization}, for which $\delta_\eps(\V r)=\delta(r-\eps)/(4\pi\eps^2)$, i.e.,  the regularized delta function is a surface delta function on a sphere of radius $\eps$. This choice is more faithful to the original cylindrical geometry than delta functions with infinite support \cite{cortez2001method, cortez2005method, maxey2001localized} or those whose support is tied to a numerical fluid grid \cite{peskin2002acta}. In particular, since each regularized singularity is a sphere, it is not hard to imagine this resulting in a cylinder of constant radius $\eps$ along a length $L$, with hemispherical caps at each end. Still, since the regularized delta function is the same in the axial and radial directions, the spheres are not rings as we would need to make a proper cylinder. Also, the RPY tensor itself is an approximation to the dynamics of spheres, since it does not include stresslet terms required to keep the spheres rigid in flow \cite{SpectralSD}, even though stresslets are of the same multipole order as torques, which are included. Because of these approximations, the intuitive picture of a cylinder composed of surface spheres is not quite the correct one, and we will require $\eps > \rc$ for our RPY formulas to match those of SBT.}

\rone{To obtain the centerline velocity of a filament made of regularized singularities, we first solve for the fluid velocity $\V{u}(x)$ and corresponding pressure $\pi \left( \V x \right)$} by\footnote{\rone{It is important to emphasize again that $\V{u}(\V x)$ and $\pi(\V x)$ are expected to be good approximations of the true fluid velocity and pressure only sufficiently far from the fiber centerline, but not $\mathcal{O}(\epsilon)$ away from it.}} solving the Stokes equations \cite[Sec.~2.1]{maxian2022hydrodynamics}
\begin{equation}
\label{eq:Stokes}
\grad\pi(\V{x})=\mu \grad^{2}\V u(\V{x})+ \int_0^L \left(\left(\V{f}(s) + \frac{\nabla}{2} \times n^\parallel(s)\Xs(s)\right)\delta_{\eps}\left(\V{x}-\V X(s)\right)\right) \, ds,
\end{equation}
and $\nabla \cdot \V u = 0$. The translational and parallel rotational velocity of the fiber centerline are then given by convolving the pointwise velocity $\V{u}(\V{x})$ and half vorticity once more with the regularized delta function, 
\begin{gather}
\label{eq:Uavg}
\V U \left(s\right)= \int \V{u}(\V{x}) \, \delta_{\eps}\left(\V X(s)-\V x\right) \, d\V{x} \\
\label{eq:OmIBdef}
\Psi^\parallel \left(s\right)=\Xs(s) \cdot \int \delta_{\eps}\left(\V X(s)-\V x\right)\frac{\grad}{2}\times\V u\left(\V x\right) \, d\V{x}.
\end{gather}
This yields a mobility that is guaranteed to be symmetric positive definite \cite[Footnote~3]{kallemov2016immersed}. Here we only consider the parallel component of rotational velocity, since, for unshearable filaments, the other components can be derived from the translational velocity $\V{U}(s)$  \cite[Sec.~2]{maxian2022hydrodynamics}.

Using superposition, the translational and rotational velocity of the fiber centerline can be defined as integrals of the fundamental \emph{regularized} singularities
\begin{align}
\label{eq:Uconv}
\V{U}(s)& = \int_0^L \left(\Mbtt\left(\V{X}(s),\V{X}(s^\prime)\right)\V{f}(s^\prime)+\Mbtr\left(\V{X}(s),\V{X}(s^\prime)\right)n^\parallel(s') \Xs(s')\right) \, ds^\prime \\ \nonumber
\Psi^\parallel(s)& = \Xs(s) \cdot \int_0^L \left(\Mbrt\left(\V{X}(s),\V{X}(s^\prime)\right)\V{f}(s^\prime)+\Mbrr\left(\V{X}(s),\V{X}(s^\prime)\right)n^\parallel(s') \Xs(s')\right) \, ds^\prime
\end{align}
where the fundamental regularized singularities $\Mbtt, \Mbtr, \Mbrt,$ and $\Mbrr$ are obtained by convolving the Stokeslet with $\delta_\eps$, with a half curl taken each time rotation comes in,
\begin{align}
\label{eq:Ukernels}
\Mbtt\left(\V{x},\V{y}\right) & =  \int \delta_{\eps}\left(\V x-\V z\right)  \int \Slet{\V{z},\V{w}}\delta_{\eps}\left(\V w-\V y\right) \, d\V{w} \, d\V{z},\\ \nonumber
\Mbtr\left(\V{x},\V{y}\right) & =  \int \delta_{\eps}\left(\V x-\V z\right)  \int \Slet{\V{z},\V{w}} \frac{\nabla_{\V{w}}}{2} \times\delta_{\eps}\left(\V w-\V y\right) \, d\V{w} \, d\V{z},\\ \nonumber
\Mbrt\left(\V{x},\V{y}\right) & =  \int \delta_{\eps}\left(\V x-\V z\right)  \frac{\nabla_{\V{z}}}{2} \times \int \Slet{\V{z},\V{w}}\delta_{\eps}\left(\V w-\V y\right) \, d\V{w} \, d\V{z}=\Mbtr^T\left(\V{y},\V{x}\right),\\ \nonumber
\Mbrr\left(\V{x},\V{y}\right) & =  \int \delta_{\eps}\left(\V x-\V z\right)  \frac{\nabla_{\V{z}}}{2} \times \int \Slet{\V{z},\V{w}} \frac{\nabla_{\V{w}}}{2} \times\delta_{\eps}\left(\V w-\V y\right) \, d\V{w} \, d\V{z}.
\end{align}

Using this approach, the four components of the mobility operator relating force and torque to angular and rotational velocity can be written \emph{separately}, giving four different integrals for the four different pieces of the velocity in\ \eqref{eq:Uconv}. Thus, we can simply do asymptotics on each of these four integrals to get an approximation to each term to some order in $\epsc$. This is in contrast to the SBTs of Section\ \ref{sec:SBT3D}, where the translational velocity on the cross section was some combination of contributions from force and torque, and we were left with trying to determine the order of each term from the desired order of $\V{U}$ and $\Psi^\parallel$ on the cross section. 

In a companion paper \cite{maxian2022hydrodynamics}, we develop efficient quadrature schemes to evaluate the integrals\ \eqref{eq:Uconv} to numerical, rather than asymptotic, precision. The RPY kernels change their form when the distance between the spheres is less than $2\eps$, i.e., when the spheres overlap, thus regularizing the singularity that occurs in the interactions between point forces. Still, the kernels are nearly singular, and in the limit $\eps/L:=\epsRS \ll 1$ it is more efficient to evaluate them asymptotically, since some of the singularities can safely be assumed to only contribute to the ``inner'' expansion. 

In this section, we therefore perform asymptotic expansions on the integrals\ \eqref{eq:Uconv} to obtain an SBT for a twisting filament. Our strategy is standard matched asymptotics and similar to the approaches of Johnson \cite{johnson} and Gotz \cite{gotz2001interactions} for SBT. Because we use regularized singularities, however, all of the integrals take place on the fiber centerline, and so the asymptotics are in one dimension, rather than on the fiber cross section (two dimensions). It therefore becomes much simpler to obtain expressions for the velocity at the endpoints (see Appendix\ \ref{sec:EPs}), where we only have to redefine the domain of integration in the inner expansion. \rone{Thus a physical endpoint regularization, which is an ingredient of any SBT-based numerical method \cite[Sec.~2.1]{maxian2021integral}, comes with our choice to use regularized singularities as an approximation to the three-dimensional fiber geometry.}

For each integral in\ \eqref{eq:Uconv}, we first compute an outer expansion by considering the region where $|s-s'|$ is $\mathcal{O}(1)$. Then in the inner expansion, we consider the part of the integrals where $|s-s'|$ is $\mathcal{O}(\eps)$. In this case, we follow \cite{gotz2001interactions} and introduce the rescaled variable
\begin{equation}
\label{eq:xidef}
\xi = \frac{s'-s}{\eps}, 
\end{equation}
so that $\xi$ is $\mathcal{O}(1)$ in the inner expansion, and the domain of $\xi$ is $[-s/\eps,(L-s)/\eps]$. With this definition, it is straightforward to integrate over $\xi$ and obtain an inner expansion, which is done in two pieces because the RPY tensor changes at $s=2\eps$ (the case of overlapping spheres).\footnote{\rone{The break up of the integral into two pieces is reminiscent of the classic approach of Lighthill \cite{lighthill1976flagellar} (see also \cite[Fig.~6]{lauga2009hydrodynamics}), who represented the flow due to a translating filament as the sum of an integral of Stokeslets (at distances greater than some intermediate length $q \gg a$) and another integral of Stokeslets and doublets (at distances smaller than $q$). The RPY approach is fundamentally different, however, because here the RPY regularization imposes the splitting of the integral at the distance $2\eps$. For translation-translation, the doublet term can still make a contribution to the flow field on this lengthscale; its precise $\mathcal{O}(1)$ contribution is computed in\ \eqref{eq:dbletin}.}} Adding the inner and outer solutions together and subtracting the common part then gives the final matched asymptotic solution. Unless otherwise specified, all expansions will be carried out to $\mathcal{O}(\epsRS)$, i.e., in this section $\approx$ denotes equality up to $\mathcal{O}(\epsRS)$.

\subsection{Translation from force}
We begin by expanding the translational velocity from force, which we define as
\begin{equation}
\label{eq:Ufdef}
\tt{\V{U}}=\int_0^L \Mbtt\left(\V{X}(s),\V{X}(s^\prime)\right)\V{f}(s^\prime) \, ds'.
\end{equation}
The following derivation is a repeat of that given in \cite[Appendix~A]{maxian2021integral}, but it is included here for completeness, and because the set of steps for the other 3 blocks of the mobility operator is identical. When we substitute the definition of $\Mbtt$ for RPY singularities from \cite[Eq.~(3.12)]{wajnryb2013generalization} into\ \eqref{eq:Ufdef}, we have the integral
\begin{align}
\label{eq:rpyint}
8\pi \mu\tt{\V{U}}\left(s\right) =  & \int_{R> 2\eps} \left(\mathbb{S}\left(\V{X}(s),\V{X}(s')\right)+\frac{2\eps^2}{3}\mathbb{D}\left(\V{X}(s),\V{X}(s')\right)\right)\V{f}(s')\, ds'\\[2 pt]
\nonumber + & \int_{R \leq 2\eps} \left(\left(\frac{4}{3\eps}-\frac{3R\left(s'\right)}{8\eps^2}\right)\M{I}+\frac{1}{8\eps^2R\left(s'\right)} \left(\V{R}\V{R}\right)\left(s'\right)\right)\V{f}\left(s'\right) \, ds',
\end{align}
where $\V{R}\left(s'\right)=\V{X}(s)-\V{X}\left(s'\right)$, and $R=\norm{\V{R}}$. The separation of the integrals captures the change in the RPY tensor when $R \leq 2\eps$. Because $s$ is an arclength parameterization, we will replace the bound $R > 2\eps$ with $|s-s'| > 2\eps$ going forward. This incurs a relative error on the order $\epsRS$, which is the same order to which we carry the asymptotic expansions.

In the outer expansion, we consider the part of the integral\ \eqref{eq:rpyint} where $|s-s'|$ is $\mathcal{O}(1)$. In this case, the doublet term in\ \eqref{eq:rpyint} is insignificant and we obtain the outer velocity by integrating the Stokeslet over the fiber centerline, 
\begin{equation}
8\pi \mu \tt{\V{U}}^{(\text{outer})}(s) = \int_{|s-s'| > 2\eps} \Slet{\V{X}(s),\V{X}(s')}\V{f}\left(s'\right) \, ds'. 
\end{equation}
The part of the kernel\ \eqref{eq:rpyint} for $R \leq 2\eps$ makes no contribution to the outer expansion since $|s-s'|$ is $\mathcal{O}(\eps)$ there.

The inner expansion of\ \eqref{eq:rpyint} is derived in a straightforward way by substituting the Taylor expansions\ \eqref{eq:innerAsymp} into the RPY velocity\ \eqref{eq:rpyint} and integrating over $\xi$. The details are given in Appendix\  \ref{sec:EPs}, with the final result that
\begin{gather}
\label{eq:Uinner}
8 \pi \mu \tt{\V{U}}^{(\text{inner})}(s) =  
\left(\M{I}+\Xs(s)\Xs(s)\right)\V{f}(s)  
\ln{\left(\frac{(L-s)s}{4\eps^2}\right)} \\ \nonumber + 
\left(\left(4-\frac{\eps^2}{3s^2}-\frac{\eps^2}{3(L-s)^2}\right)\M{I} + \left(\frac{\eps^2}{s^2}+\frac{\eps^2}{(L-s)^2}\right) \Xs(s)\Xs(s)\right)\V{f}(s),
\end{gather}
in the fiber interior. The complete formulas including endpoint modifications are given in\ \eqref{eq:UinnerA}; the $\mathcal{O}\left(\eps^2\right)$ terms are included in\ \eqref{eq:Uinner} because they are necessary to give $\mathcal{C}^2$ continuity with the endpoint formulas.

The common part is the outer velocity written in terms of the inner variables, 
\begin{equation}
8\pi \mu \tt{\V{U}}^{(\text{common})}(s) = \int_{|s-s'| > 2\eps} \left(\frac{\M{I}+\Xs(s)\Xs(s)}{|s-s'|}\right)\V{f}(s) \, ds'.
\end{equation}

The total velocity is then the sum of the inner and outer expansions, with the common part subtracted, 
\begin{gather}
\tt{\V{U}}(s) = \tt{\V{U}}^{(\text{inner})}(s) + \tt{\V{U}}^{(\text{outer})}(s)-\tt{\V{U}}^{(\text{common})}(s). 
\end{gather}
This can be written as 
\begin{gather}
\label{eq:totvelnofp}
\tt{\V{U}}(s) = \tt{\V{U}}^{(\text{inner})}(s) + \EPMI\int_{|s-s'|> 2\eps} \left(\Slet{\V{X}(s),\V{X}(s')} \V{f}\left(s'\right) -   \left(\frac{\M{I}+\Xs(s)\Xs(s)}{|s-s'|}\right)\V{f}(s)\right) \, ds', 
\end{gather}
where $\tt{\V{U}}^{(\text{inner})}$ is defined in\ \eqref{eq:UinnerA}.
The velocity\ \eqref{eq:totvelnofp} has a form similar to that of SBT, but with different bounds on the integral. Unlike SBT, the integral\ \eqref{eq:totvelnofp} is not a finite part integral, but a nearly singular integral that makes sense for each term separately. Numerical evidence suggests that, unlike traditional SBT, our asymptotic formula\ \eqref{eq:totvelnofp} keeps the eigenvalues of the mobility above zero \cite[Sec~4.4.1]{maxian2022hydrodynamics}.

To compare with SBT, however, it is advantageous to observe that the integrand in\ \eqref{eq:totvelnofp} is $\mathcal{O}(\epsRS)$ when $R \leq 2\eps$, and so we can add the excluded part back in without changing the asymptotic accuracy. This gives a velocity of the exact same form as SBT,
\begin{gather}
\label{eq:totvelfp}
8 \pi \mu \tt{\V{U}}(s) = \left(\M{I}+\Xs(s)\Xs(s)\right)\V{f}(s)  
\ln{\left(\frac{(L-s)s}{4\eps^2}\right)} +4\V{f}(s)\\[4 pt] \nonumber
 + \int_{0}^L \left(\Slet{\V{X}(s),\V{X}(s')} \V{f}\left(s'\right) -  \left(\frac{\M{I}+\Xs(s)\Xs(s)}{|s-s'|}\right)\V{f}(s)\right) \, ds', 
\end{gather}
in the fiber interior, where we have dropped $\mathcal{O}(\eps^2)$ terms from\ \eqref{eq:Uinner}. The comparison of this formula to translational SBT will be given in Section\ \ref{sec:compareRPSB}.


\subsection{Translation from torque \label{sec:rpyttorq}}
To complete our formulation for translational velocity, we next consider an asymptotic expansion of the velocity $\tr{\V{U}}(s)$ due to a torque density $n^\parallel(s)$ on the fiber centerline,
\begin{equation}
\label{eq:Urdef}
\tr{\V{U}}(s)=\int_0^L \Mbtr\left(\V{X}(s),\V{X}(s^\prime)\right)n^\parallel(s') \Xs(s') \, ds'.
\end{equation}
When we substitute the definition of $\Mbtr$ for RPY singularities from \cite[Eq.~(3.16)]{wajnryb2013generalization} into\ \eqref{eq:Urdef}, we have the integral
\begin{gather}
\label{eq:tfromt}
8\pi \mu \tr{\V{U}}(s) = \int_{R > 2\eps} \frac{\Xs(s') \times \V{R}(s')}{R(s')^3}n^\parallel(s') \, ds'\\[4 pt] \nonumber + 
\frac{1}{2\eps^2}\int_{R \leq 2\eps} \left(\frac{1}{\eps}-\frac{3R(s')}{8\eps^2}\right)\left(\Xs(s') \times \V{R}(s')\right)n^\parallel(s') ds',
\end{gather}
where $\V{R}(s')=\V{X}(s)-\V{X}(s')$. The first term in\ \eqref{eq:tfromt} is the Rotlet, while the second term is the RPY tensor when $R \leq 2\eps$. As before, we will use $R \approx |s-s'|$ to modify the bounds on the integrals.

In the outer expansion, we consider the part of the integral\ \eqref{eq:tfromt} where $|s'-s|$ is $\mathcal{O}(1)$. In this case the correction term is insignificant and we get the outer velocity
\begin{equation}
\label{eq:outerTT}
8\pi \mu \tr{\V{U}}^{\text{(outer)}}(s) =  \int_{|s-s'| > 2\eps} \frac{\Xs(s') \times \V{R}(s')}{R(s')^3}n^\parallel(s') \, ds'.
\end{equation}

In the inner expansion, we use the inner asymptotics\ \eqref{eq:innerAsymp} to write the inner expansion as
\begin{align}
\label{eq:UinTT}
8\pi\mu \tr{\V{U}}^{\text{(inner)}}(s) &= \frac{1}{2}\left(\Xs(s) \times \ds{\Xs}(s)\right)\left(\ln{\left(\frac{s(L-s)}{4\eps^2}\right)}+\frac{7}{6}\right)
\end{align}
in the fiber interior, with the corresponding endpoint formula given in\ \eqref{eq:UinTTA}. The details of this inner expansion can be found in Appendix\ \ref{sec:EPs}.

The common part for trans-rot coupling is the inner expansion written in terms of the outer variables,
\begin{equation}
\label{eq:commonTT}
8\pi\mu \tr{\V{U}}^{\text{(common)}}(s) =\int_{|s-s'| > 2\eps} \left(\frac{\Xs(s)  \times \ds{\Xs}(s)}{2|s-s'|}\right)n^\parallel(s) \, ds'.
\end{equation}

To $\mathcal{O}(\epsRS)$, the velocity due to torque $\V{n}(s)=n(s)\Xs(s)$ on the fiber centerline is given by
\begin{gather}
\tr{\V{U}}(s)= \tr{\V{U}}^{\text{(inner)}}(s)+\tr{\V{U}}^{\text{(outer)}}(s)-\tr{\V{U}}^{\text{(common)}}(s)\\[4 pt]
\label{eq:match1}
 \tr{\V{U}}(s) = \tr{\V{U}}^{\text{(inner)}}(s)+ \EPMI \int_{R > 2\eps}  \left(\frac{\Xs(s') \times \V{R}(s')}{R(s')^3}n^\parallel(s')- \frac{\Xs(s)  \times \ds{\Xs}(s)}{2|s-s'|} n^\parallel(s)\right) \, ds' , 
\end{gather}
We can make this velocity\ \eqref{eq:match1} more SBT-like by changing the bounds on the integral. Because the outer and common part match to $\mathcal{O}(\epsRS)$ when $R \leq 2\eps$, we can change the limit of integration in the integral to obtain a finite part integral, which gives the final result
\begin{gather}
\label{eq:finalTT}
8\pi \mu \tr{\V{U}}(s) =\frac{1}{2}\left(\Xs(s) \times \ds{\Xs}(s)\right)\left(\ln{\left(\frac{s(L-s)}{4\eps^2}\right)}+\frac{7}{6}\right)\\[4 pt] \nonumber+ \int_{0}^L  \left(\left(\frac{\Xs(s') \times \V{R}(s')}{R(s')^3}\right)n^\parallel(s')-\left(\frac{\Xs(s) \times \ds\Xs(s)}{2|s-s'|}\right) n^\parallel(s) \right) \, ds' ,
\end{gather}
for the velocity in the fiber interior. This is of exactly the same form as the first line of the singularity representation\ \eqref{eq:uOmEx}, but without the additional angle-dependent terms in the second line of\ \eqref{eq:uOmEx}. 



\subsection{Rotation from force}
We next calculate the parallel rotational velocity $\rt{\Psi}^\parallel(s)$ due to the force density $\V{f}(s)$ on the fiber centerline, 
\begin{equation}
\label{eq:Psifdef}
\rt{\Psi}^\parallel(s)=\Xs(s) \cdot \int_0^L \Mbrt\left(\V{X}(s),\V{X}(s^\prime)\right)\V{f}(s') \, ds'.
\end{equation}
When we substitute the definition of $\Mbtr$ from \cite[Eq.~(3.16)]{wajnryb2013generalization} into\ \eqref{eq:Psifdef}, we have
\begin{gather}
\label{eq:rfromf}
8\pi \mu \rt{\Psi}^\parallel(s)=\int_{R > 2\eps} \frac{\left(\V{f}(s') \times \V{R}(s')\right)}{R(s')^3}\cdot \Xs(s) \, ds'\\[4 pt] \nonumber + 
\frac{1}{2\eps^2}\int_{R < 2\eps} \left(\frac{1}{\eps}-\frac{3R(s')}{8\eps^2}\right)\left(\V{f}(s') \times \V{R}(s')\right)\cdot \Xs(s)\, ds'.
\end{gather}
After using the triple cross product identity to rewrite $\left(\V{f}(s') \times \V{R}(s')\right) \cdot \Xs(s)=\left(\V{R}(s') \times \Xs(s)\right) \cdot \V{f}(s')$, the symmetry of the kernels $\Mbrt$ and $\Mbtr$ makes the asymptotics in the inner, outer, and common expansions much the same as the previous section. The final result is that 
\begin{gather}
\label{eq:matchRFp}
8\pi \mu \rt{\Psi}^\parallel(s) = \frac{1}{2}\left(\ln{\left(\dfrac{s(L-s)}{4\eps^2}\right)}+ \dfrac{7}{6}\right)\left(\Xs(s) \times \ds{\Xs}(s)\right) \cdot \V{f}(s)+\rt{\Psi}^\text{(FP)}(s)\\[2 pt] 
\label{eq:OmegaFP}
\rt{\Psi}^\text{(FP)}(s) =  \int_0^L  \left(\frac{\V{R}(s') \times \Xs(s)}{R(s')^3} \cdot \V{f}(s')-\frac{\Xs(s) \times \ds{\Xs}(s)}{2|s-s'|}\cdot \V{f}(s)\right)\, ds'.
\end{gather}
in the fiber interior. To obtain a formula accurate to $\mathcal{O}(\epsRS)$ up to and including the fiber endpoints, we replace the first line in\ \eqref{eq:matchRFp} with $\rt{\Psi}^{\parallel, \text{inner}}$ defined in\ \eqref{eq:matchRFApp}.

Using the endpoint formulas in Appendix\ \ref{sec:EPs}, it is straightforward to show the symmetry (self-adjointness in $L^2$) of the mobility operator,
\begin{equation}
\int_0^L \rt{\Psi}^\parallel(s) n^\parallel(s) \, ds = \int_0^L \tr{\V{U}}(s) \cdot \V{f}(s) \, ds.
\end{equation}
The only nontrivial part of this calculation is the rotlet term in the finite part integrals, in which the order of integration in $s$ and $s'$ must be swapped. 


\subsection{Rotation from torque}
The final asymptotic expansion we need is the rotational velocity $\V{\Psi}(s)$ due to the torque $n^\parallel(s)$. Because the doublet singularity has already been expanded in the translational case in\ \eqref{eq:dbletin}, it will be convenient to work with a full vector rotational velocity and torque
\begin{equation}
\label{eq:Psirdef}
\rr{\V{\Psi}}(s)=\int_0^L \Mbrr\left(\V{X}(s),\V{X}(s^\prime)\right)\V{n}(s')  \, ds',
\end{equation}
and specialize to the case of parallel velocity and torque later. We substitute the definition of $\Mbrr$ from \cite[Eq.~(3.14)]{wajnryb2013generalization} into\ \eqref{eq:Psirdef} to obtain the vector rotational velocity,
\begin{gather}
\label{eq:rfromt}
8\pi \mu \rr{\V{\Psi}}(s) = -\frac{1}{2}\int_{R > 2\eps} \mathbb{D}\left(\V{X}(s),\V{X}(s')\right) \V{n}(s')  \, ds'\\[2 pt] 
\nonumber
+\frac{1}{\eps^3}\int_{R \leq 2\eps} \left(\left(1-\frac{27R(s')}{32\eps}+\frac{5 R(s')^3}{64 \eps^3}\right)\M{I}+\left(\frac{9}{32\eps R(s')}-\frac{3R(s')}{64\eps^3}\right)\left(\V{R}\V{R}\right)(s')\right)\V{n}(s') \, ds'.
\end{gather}
The first term in\ \eqref{eq:rfromt} is the doublet, while the second term is the RPY tensor for $R \leq 2\eps$.

In the outer expansion, we consider the part of the integral\ \eqref{eq:rfromt} where $|s'-s|$ is $\mathcal{O}(1)$,
\begin{equation}
\label{eq:outerTT}
8\pi \mu  \rr{\V{\Psi}}^{\text{(outer)}}(s) =  -\frac{1}{2}\int_{|s-s'| > 2\eps}  \mathbb{D}\left(\V{X}(s),\V{X}(s')\right) \V{n}(s')   \, ds'.
\end{equation}
Importantly, the outer expansion (and the common expansion) is $\mathcal{O}(1)$. The inner expansion will be $\mathcal{O}\left(\epsRS^{-2}\right)$, so we will wind up dropping the entire outer expansion.

In Appendix\ \ref{sec:EPs}, we use the inner asymptotics\ \eqref{eq:innerAsymp} to derive the inner angular velocity
\begin{gather}
\label{eq:EPRR}
8\pi \mu\rr{\V{\Psi}}^{\text{(inner)}}(s)=\left(p_I(s) \M{I}+p_\tau(s) \M{\Xs}(s)\Xs(s)\right)\V{n}(s), \qquad \text{where} \\[4 pt] \nonumber
p_I (s)= \dfrac{1}{\eps^2} \left(\dfrac{9}{8}+\dfrac{\eps^2}{4}\left(\dfrac{1}{s^2}+\dfrac{1}{(L-s)^2}\right)\right)\\ \nonumber
p_\tau(s) = \dfrac{1}{\eps^2}\left(\dfrac{9}{8}-\dfrac{3\eps^2}{4}\left(\dfrac{1}{s^2}+\dfrac{1}{(L-s)^2}\right)\right),
\end{gather}
in the fiber interior. The functions $p_I$ and $p_\tau$ are different at the fiber endpoints, and are given in\ \eqref{eq:EPRRA}. As in the translation case, the $\mathcal{O}\left(\eps^2\right)$ terms in\ \eqref{eq:EPRR} are negligible in the fiber interior, but are necessary to give $\mathcal{C}^2$ continuity with the endpoint formulas in\ \eqref{eq:EPRRA}.

Because the outer expansion is $\mathcal{O}(\epsRS^2)$ smaller than the inner expansion, the total asymptotic velocity is just given by the inner expansion, 
\begin{equation}
\label{eq:cylrot}
\rr{\V{\Psi}}= \rr{\V{\Psi}}^{\text{(inner)}}.
\end{equation}
Specializing now to the case of parallel angular velocity from parallel torque, $\V{n}=n^\parallel \Xs$, and $\rr{\Psi}^\parallel = \rr{\V{\Psi}} \cdot \Xs$, and dropping the negligible terms in the fiber interior from\ \eqref{eq:EPRR}, we obtain the simple formula
\begin{equation}
\label{eq:RRFinal}
8 \pi \mu \rr{\Psi}^\parallel(s)  = \frac{9n^\parallel(s)}{4\eps^2}
\end{equation}
in the fiber interior.

\subsection{Comparison to Keller-Rubinow-Johnson \label{sec:compareRPSB}}
\rone{We recall that the goal of our regularized singularity approach is to find $\eps$ such that the RPY-based filament centerline velocity matches that of SBT. With this in mind, let us compare our RPY-based slender body theory to the more rigorous three-dimensional SBT results we have from Section\ \ref{sec:SBT3D}. }We first compare the theories when we have a well-defined SBT, i.e., for translation-translation and rotation-rotation. Then, we use the form of the translation-rotation RPY coupling and our results in Section\ \ref{sec:SBT3D} to postulate an SBT equation for trans-rot coupling with a single unknown parameter. The parameter will be determined in the next section.

\subsubsection{Trans-trans and rot-rot \label{sec:rpysbttt}}
All three-dimensional slender body theories agree that the translational velocity in the case when $\Psi^\parallel=0$ is given by
\begin{gather}
\label{eq:TransSBT}
8 \pi \mu \tt{\V{U}}=\ln{\left(\frac{4s(L-s)}{\rc ^2}\right)} \left(\M{I}+\Xs \Xs\right) \V{f} + \left(\M{I}-3\Xs \Xs\right) \V{f}+\tt{\V{U}}^{\text{(FP)}}+\mathcal{O}(\epsilon),
\end{gather}
while our RPY-based theory\ \eqref{eq:totvelfp} gives
\begin{gather}
8 \pi \mu \tt{\V{U}}=\ln{\left(\frac{s(L-s)}{4\eps ^2}\right)} \left(\M{I}+\Xs \Xs\right) \V{f} + 4\V{f}+\tt{\V{U}}^{\text{(FP)}}+\mathcal{O}(\epsilon).
\end{gather}
For the case of rotation from parallel torque, all theories of Section\ \ref{sec:SBT3D} agree that
\begin{equation}
8 \pi \mu \rr{\Psi}^\parallel=\frac{2 n^\parallel }{\rc^2},
\end{equation}
while our RPY-based theory gives\ \eqref{eq:RRFinal}. 

While we can set $\eps$ to match the two translational theories or the two rotational theories, there is no choice of $\eps$ that will match both. There are thus two possible candidates for $\eps$. If we take 
\begin{equation}
\eps=\eps_\text{tt} = \rc \frac{e^{3/2}}{4} \approx 1.12 \rc,
\end{equation}
then the two translational theories match \cite[Appendix~A]{maxian2021integral}, but for rotation we obtain
\begin{gather}
8\pi\mu\rr{{\Psi}}^\parallel=\frac{9n^\parallel(s)}{4\eps_\text{tt}^2}=\frac{36n^\parallel(s)}{e^3 \rc^2} \approx 1.79 \frac{n^\parallel}{\rc^2},
\end{gather}
which is a 10\% error from the known formula for a cylinder\ \eqref{eq:cylrot}. 

On the other hand, if we take 
\begin{equation}
\eps=\eps_\text{rr} = \sqrt{9/8} \rc \approx 1.06 \rc,
\end{equation}
the rotational mobilities match, and the resulting RPY translational mobility becomes
\begin{align}
\nonumber
8 \pi \mu \V{U}&=\ln{\left(\frac{s(L-s)}{4\eps_\text{rr} ^2}\right)} \left(\M{I}+\Xs \Xs\right) \V{f} + 4\V{f}+\tt{\V{U}}^{\text{(FP)}} \\ \nonumber
& \approx  \ln{\left(\frac{4s(L-s)}{\rc ^2}\right)}\left(\M{I}+\Xs \Xs\right) \V{f} + \left(1.1\M{I}-2.9\Xs \Xs\right) \V{f} + \tt{\V{U}}^{\text{(FP)}},
\end{align}
which gives a small $\mathcal{O}(1)$ error relative to translational SBT\ \eqref{eq:TransSBT}. \rthr{The need to choose a different regularized singularity radius for the two different mobility components shows that,  despite the intuitive picture of making a cylinder out of spheres, our choice of spherically-symmetric regularized singularities is insufficient to exactly model a cylinder. It is clear, however, that the RPY approximation captures the correct form of all the relevant terms, but with $\mathcal{O}(1)$ error in the coefficients.}

\subsubsection{Trans-rot coupling}
The more compelling question is whether we can use our findings to gain insights into the unknown $\mathcal{O}(1)$ coefficient in translation-rotation SBT. Our RPY-based result is \rone{(rearranging\ \eqref{eq:finalTT} using properties of logs and rounding the $\mathcal{O}(1)$ coefficient to two decimal places)}
\begin{gather}
\label{eq:RotTransRPY}
\rone{8\pi \mu \tr{\V{U}} \approx \left(\ln{\left(\frac{4 s(L-s)}{\eps^2}\right)}-1.61\right)\left(\Xs \times \ds \Xs\right)\frac{n^\parallel}{2}+\rt{\V{U}}^{\text{(FP)}},}
\end{gather}
with the corresponding symmetric result for $\rt{\Psi}^\parallel$ in\ \eqref{eq:matchRFp}. Since Section\ \ref{sec:rpysbttt} shows that the RPY models capture the essential physics of the interaction of the fiber with the fluid, but with different coefficients, we can postulate that the rotation-translation coupling in a true tubular geometry is also of the form\ \eqref{eq:RotTransRPY}, but with different coefficients. Furthermore, since there is an $\mathcal{O}(1)$ error in the boundary condition\ \eqref{eq:UMatch} in the result\ \eqref{eq:uOmEx} of Johnson, we can expect that any modifications to\ \eqref{eq:uOmEx} will be $\mathcal{O}(1)$ (not $\log{\epsc}$). Combining these two assumptions, we postulate a rotation-translation coupling term of the form 
\begin{equation}
\label{eq:kRT2}
8\pi \mu \tr{\V{U}} =  \left(\ln{\left(\frac{4s(L-s)}{ \rc^2}\right)}-k\right) \left(\Xs \times \ds \Xs\right)\frac{n^\parallel}{2}+\tr{\V{U}}^{\text{(FP)}}+\mathcal{O}(\epsc)
\end{equation}
for some unknown constant $k$, with the corresponding symmetric form for $\rt{\Psi}^\parallel$. In the next section, we fit the constant $k$ to three dimensional boundary integral results. This gives us an empirical three-dimensional SBT for rot-trans coupling to which we can compare the RPY theory. \rone{It is important to note that our work does not rigorously justify\ \eqref{eq:kRT2} for a specific three-dimensional fiber geometry.}


\section{Comparing to three-dimensional boundary integral \label{sec:BI}}
In this section, we compare our slender body theories of Sections\ \ref{sec:SBT3D} and\ \ref{sec:RPYSBT} to a three-dimensional boundary integral (BI) calculation. Because there is one uncertain coefficient, the $\mathcal{O}(1)$ coefficient $k$ in the rot-trans coupling term\ \eqref{eq:kRT2}, the goal of these calculations is to find a $k$ that can predict the full BI results to $\mathcal{O}(\epsc)$ accuracy. Because of this, we can fix the fiber geometry, vary the aspect ratio $\epsc$, and check that the error in SBT decreases with $\epsc$. It will be convenient\footnote{\rone{While the result\ \eqref{eq:RotTransRPY} was derived by assuming a constant regularization radius $\eps$, it is unchanged to leading order when considering a spatially-varying radius function $\eps(s)$, for which the polydisperse RPY kernel must be used \cite{RPY_Shear_Wall}. The choice of an ellipsoidally-decaying radius function removes any singular behavior of\ \eqref{eq:kRT2} near the endpoints and speeds up the convergence of the boundary integral method (relative to a cylinder with spherical caps, for example).} 
} to fix $L=2$ and work on $s \in [-1,1]$ with an ellipsoidally-decaying radius function
 \begin{equation}
\label{eq:rhoel}
\rho(s) = \sqrt{1-s^2}.
\end{equation}
The radius at each point $s$ is $\rc \rho(s)$, where we will vary $\rc$ to study the limit $\epsc=\rc/L \rightarrow 0$. 

As discussed in Section\ \ref{sec:KL}, under the assumption that the fiber is an infinitely thin membrane containing fluid of viscosity equal to that of the exterior fluid, the force density multiplied by the surface area element on the tubular boundary can be obtained by solving the integral equation
\begin{gather}
\label{eq:SingLayer}
8 \pi \mu \Uhat(s,\theta)  = \int_{-1}^1 \int_{0}^{2\pi} \mathbb{S}\left(\Xhat(s,\theta),\Xhat \left(s',\theta'\right)\right) \fhat \left(s',\theta'\right) \, d\theta' ds' ,
\end{gather}
for $\fhat(s,\theta)$, where $\Xhat$ and $\widehat{\V{U}}$ are defined in\ \eqref{eq:Xhat} and\ \eqref{eq:Uwant}. 

In order to compare the BI output to SBT, we use the following sequence of steps:
\begin{enumerate}
\item Given the surface velocity $\Uhat(s,\theta)= \V{U}(s) + \rc \Psi^\parallel(s) \left(\Xs(s) \times \widehat{\V{r}}(s,\theta)\right)$, solve\ \eqref{eq:SingLayer} using the numerical method of Section\ \ref{sec:BInumer}  to obtain the surface force density $\fhat(s,\theta)$.
\item Compute the centerline force $\V{f}(s)$ and parallel torque $n^\parallel(s)$ from $\fhat(s,\theta)$ via\ \eqref{eq:fTrac} and\ \eqref{eq:nparTrac}.
\item Compute the SBT velocity induced from the BI force as
\begin{gather}
\label{eq:SBTtrans}
8 \pi \mu\Usb=  \ln{\left(\epsc^{-2}\right)} \left(\M{I}+\Xs \Xs\right) \V{f} + \left(\M{I}-3\Xs \Xs\right) \V{f}+\tt{\V{U}}^{\text{(FP)}} \\ \nonumber
+ \left(\ln{\left(\epsc^{-2}\right)}-k\right) \left(\Xs \times \ds \Xs\right)\frac{n^\parallel}{2}+\tr{\V{U}}^{\text{(FP)}}, \\
\label{eq:kRT}
8 \pi \mu\Omsb=  \frac{2n^\parallel}{\rc^2 \rho^2}+\frac{1}{2}\left(\ln{\left(\epsc^{-2}\right)}-k\right)\left(\Xs \times \ds{\Xs}\right) \cdot \V{f}+\rt{\Psi}^\text{(FP)},
\end{gather}
where the finite part integrals are defined in\ \eqref{eq:UfFP},\ \eqref{eq:UnFP}, and\ \eqref{eq:OmegaFP}, and we have used the local drag coefficients for ellipsoidally-decaying radius\ \eqref{eq:rhoel}. 
\item Calculate the $L^2$ norm of the error $\Usb-\V{U}$ and $\Omsb - \Psi^\parallel$, and plot it as a function of $\epsc$.
\end{enumerate}
Note that we apply the forward SBT operator, which gives velocity from force/torque, in\ \eqref{eq:SBTtrans} and\ \eqref{eq:kRT}. This avoids inverting the slender body mobility, which is an ill-posed problem \cite{gotz2001interactions, ts04}. To evaluate the finite part integrals, we use the singular quadrature scheme developed in \cite{tornquad, maxian2021integral, maxian2022hydrodynamics}.

We have verified that the error in $\Usb$ and $\Omsb$ decreases with $\epsc$ in cases when rot-trans coupling is negligible (i.e., cases when $\Omsb=0$ or the fiber is straight). Having done this, our focus will be on the case of curved filaments with parallel rotational velocity $\Psi^\parallel \sim 1/\epsc^2$, so that the magnitude of the parallel torque $n^\parallel$ is unchanged to leading order as we change $\epsc$. In this case, which is the one that exposed the flaws in the SBTs of Section\ \ref{sec:SBT3D}, the rot-trans coupling term in\ \eqref{eq:SBTtrans} is important, while the one in\ \eqref{eq:kRT} is negligible. Our goal is to find a $k$ in\ \eqref{eq:SBTtrans} such that the formulas are accurate to $\mathcal{O}(\epsc)$; that is, the relative errors in both $\norm{ \Usb-\V{U}}$ and $\Omsb - \Psi^\parallel$ are $\mathcal{O}(\epsc)$.

\subsection{Numerical method \label{sec:BInumer}}
To solve\ \eqref{eq:SingLayer}, we need a way to integrate the singularity at $\left(s',\theta'\right) = \left(s,\theta\right)$. We follow the idea of Batchelor \cite{batchelor1970slender} and Koens \cite{koens2022tubular} in adding and subtracting the flow from a translating, rigid prolate ellipsoid, which has constant $\fhat$ along its surface. The geometry of the prolate ellipsoid is chosen to exactly cancel the singularity in\ \eqref{eq:SingLayer}. Specifically, we introduce an ellipsoid with geometry
\begin{equation}
\label{eq:Xell}
\Xhat_e\left(s',\theta',s,\theta\right) = \V{X}(s) + \ell(s,\theta) \left(s'-s_c(s)\right) \Xs(s) + a_e(s)  \rhat(s,\theta') \sqrt{1-\left(s'\right)^2}.
\end{equation}
This ellipsoid has half axis length $\ell$ and radius $a_e$. The point $s_c(s)$ is the coordinate on the ellipsoid at which we want to cancel the singularity at $\Xhat\left(s,\theta\right)$. Following Koens \cite{koens2022tubular} again, we cancel the singularity by imposing the conditions
\begin{gather}
\label{eq:matchSing}
\Xhat_e \left(s_c,\theta',s,\theta\right) =\Xhat \left(s,\theta\right) \\ \nonumber \partial_{s'} \Xhat_e \left(s_c,\theta,s,\theta\right)= \ds{\Xhat} (s,\theta) \qquad \partial_{\theta'} \Xhat_e \left(s_c,\theta,s,\theta\right)=\partial_\theta \Xhat (s,\theta).
\end{gather}
Substituting the definition of $\Xhat$ from\ \eqref{eq:Xhat} and $\Xhat_e$ from\ \eqref{eq:Xell}, the matching conditions\ \eqref{eq:matchSing} give three equations for the unknown parameters $\ell, a_e$, and $s_c$ in\ \eqref{eq:Xell}, 
\begin{gather}
\label{eq:aesc1}
\rc \rho = a_e \sqrt{1-s_c^2}\\
\label{eq:ell}
 1+\rc \rho \left(\Xs \cdot \ds \rhat \right) = \ell \\
 \rc \ds{\rho} = -\frac{a_e s_c}{\sqrt{1-s_c^2}}.
\end{gather}
The first of these equations matches the positions and automatically matches the $\theta$ derivatives, while the second two are obtained by projecting the $s$ derivative equation onto the $\Xs$ and $\rhat$ directions, respectively. To solve for $s_c, a_e$, and $\ell$, we first observe that\ \eqref{eq:ell} gives $\ell$ in terms of the fiber geometry, and so we are left with a $2 \times 2$ nonlinear system for $a_e$ and $s_c$. This system has two solutions, only one of which gives $s_c \in [-1,1]$,
\begin{gather}
s_c = \frac{\rho - \sqrt{\rho^2 + 4 \left(\ds \rho\right)^2}}{2\rho},
\end{gather}
with $a_e$ determined from $s_c$ using\ \eqref{eq:aesc1}. When the fiber is a straight ellipsoid, $\rho(s) = \sqrt{1-s^2}$ and $\ds \rhat=0$, so $a_e=\rc$ and $s_c=s$, and therefore the effective ellipsoid reduces to the ellipsoid itself.

Because the traction multiplied by the area element on a translating prolate ellipsoid is constant, the singularity subtraction scheme for\ \eqref{eq:SingLayer} can now be written as \cite[Eq.~5]{koens2022tubular}
\begin{gather}
\label{eq:SingLayerS}
\Uhat(s,\theta)  = \M{M}_e(s,\theta) \V{f}(s,\theta)+ \EPMI \int_{-1}^1  \int_{0}^{2\pi} \Bigg{(}\mathbb{S}\left(\Xhat(s,\theta),\Xhat \left(s',\theta'\right)\right) \V{f}\left(s',\theta'\right)\\ \nonumber -\mathbb{S}\left(\Xhat_e(s_c,\theta,s,\theta),\Xhat_e\left(s',\theta',s,\theta \right)\right) \V{f}\left(s,\theta\right)\Bigg{)} \, d\theta' ds' ,
\end{gather}
where the matrix $ \M{M}_e(s,\theta)$ can be determined from analytical formulas for the resistance on a prolate ellipsoid. In particular, defining the eccentricity
\begin{equation}
e = \sqrt{1-\frac{a_e^2}{\ell^2}}, \qquad q = \log{\left(\frac{1+e}{1-e}\right)},
\end{equation}
the resistance on a translating prolate ellipsoid is  \cite[Eqs.~(44--45)]{keaveny2011applying} (see also \cite{padrino2020comment})
\begin{gather}
\xi_\parallel = \frac{16 \pi \ell e^3 \mu}{(1+e^2)q-2e} \qquad \xi_\perp = \frac{32 \pi \ell e^3 \mu}{(3e^2-1)q+2e}
\end{gather}
These give the relationship $F_\parallel = U_\parallel \xi_\parallel$ and $F_\perp=U_\perp \xi_\perp$. To obtain the traction times area element on the surface, we divide the resistance by $4 \pi$ and invert the result, thus obtaining an expression for the matrix $\M{M}_e$, 
\begin{gather}
\M{M}_e  = \frac{4 \pi}{\xi_\parallel}\Xs(s)\Xs(s) + \frac{4\pi}{\xi_\perp}\left(\M{I}-\Xs(s) \Xs(s)\right).
\end{gather}
Because the singularity in\ \eqref{eq:SingLayerS} is canceled, we can now discretize the integral by standard quadrature, with the singular point given a weight of zero (i.e., skipped). We will use $N_\theta$ uniformly spaced points in $\theta$ and $N$ Clenshaw-Curtis quadrature points in $s$. By assigning the singular point a weight of zero, we reduce the accuracy of this quadrature scheme, which is spectrally accurate for smooth integrands, to first order, as demonstrated in Appendix\ \ref{sec:BIConv}. Our Matlab implementation of this quadrature scheme is available at \url{https://github.com/stochasticHydroTools/SlenderBody/tree/master/Matlab/NumericalSBT/SingleLayer}.

\subsection{Rotating curved fiber \label{sec:FindK}}
To extract the unknown coefficient $k$ in the trans-rot coupling term in\ \eqref{eq:SBTtrans}, we consider the half-turn helix
\begin{equation}
\label{eq:Xhelix}
\V{X}(s) = \frac{1}{\sqrt{2}}\left(\frac{1}{\pi}\cos{\left(\pi s\right)},\frac{1}{\pi}\sin{\left(\pi s\right)}, s\right),
\end{equation}
which spins with angular velocity $\Psi^\parallel  \equiv 1/\rc^2$ and without moving, $\V{U}=0$\ in\ \eqref{eq:Uwant}. Because the fiber is not straight, the torque required to generate $\Psi^\parallel$ will induce a translational velocity, and a nonzero average force $\V{f}$ must cancel this translational velocity to satisfy the boundary condition~\eqref{eq:UMatch}. This means that the translation-rotation coupling term in\ \eqref{eq:SBTtrans} will play a key role in the accuracy of our SBT. Because of the large torque, the rotational velocity in\ \eqref{eq:kRT} is dominated by the rotation-rotation mobility, and so to study translation-rotation coupling we will focus only on the translational velocity\ \eqref{eq:SBTtrans}.

\begin{figure}
\centering
\includegraphics[width=0.35\textwidth]{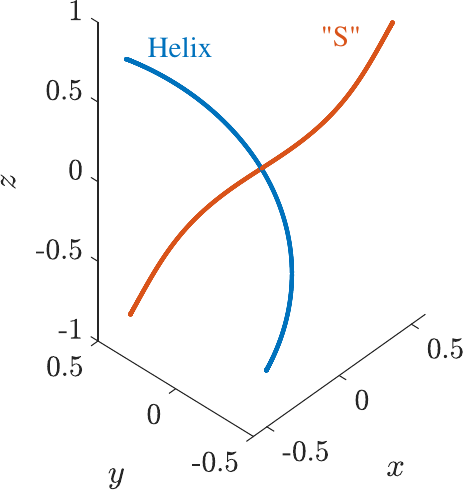}
\includegraphics[width=0.6\textwidth]{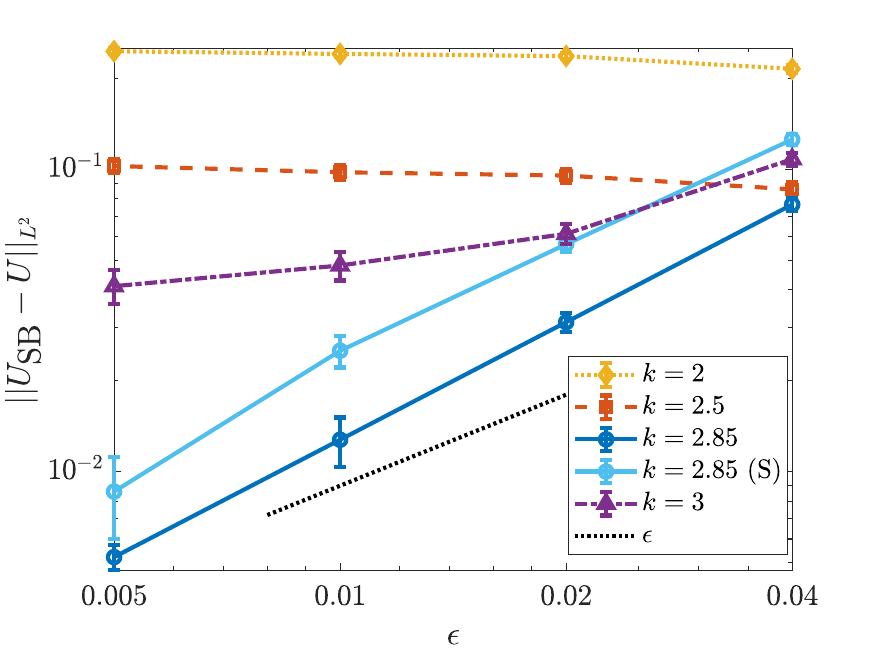}
\caption{\label{fig:TransRotErk}Error in the rot-trans SBT relative to the boundary integral calculation for various $\epsc$ and $k$ \rone{and two different fiber shapes. The fiber shapes we consider, shifted so that their midpoints coincide, are shown at left.} For each fiber, we prescribe a rotational velocity $\Psi^\parallel \equiv 1/\rc^2$ with translational velocity $\V{U}=\V 0$. We then solve the boundary integral equation\ \eqref{eq:SingLayer} for the surface traction, use\ \eqref{eq:fTrac} and\ \eqref{eq:nparTrac} to obtain the centerline force density and parallel torque density, and then\ \eqref{eq:SBTtrans} to obtain an SBT velocity $\Usb$ from these densities. We plot the $L^2$ norm of $\Usb$ (which should be zero) for several different values of $k$. For $k \neq 2.85$, \rone{results are shown only for the half-helix\ \eqref{eq:Xhelix}, and the error is clearly $\mathcal{O}(1)$ with respect to $\epsc$. For $k=2.85$, the error decreases with $\epsc$ at a rate of roughly $\epsc^1$, independent of the fiber shape considered}. The data points here come from the most refined discretization in Fig.\ \ref{fig:TransRotConv}, while the error bars are the difference between the most refined and second-most refined calculation. }
\end{figure}

In Appendix\ \ref{sec:BIConv}, we verify first-order accuracy in obtaining the cross sectional force $\V{f}(s)$ in\ \eqref{eq:fTrac} from the traction jump $\fhat(s,\theta)$ in\ \eqref{eq:SingLayerS} as we refine $N_\theta$ and $N$ \emph{together}. We also show how the errors $\norm{\Usb-\V{U}}_{L^2}$, which are obtained from the traction jump via\ \eqref{eq:fTrac} and then\ \eqref{eq:SBTtrans}, change as we refine $N_\theta$ and $N$ for three different values of $k$ ranging from $k= 2$ (which corresponds to throwing out the second line of the rotlet line integral\ \eqref{eq:uOmEx}), to $k=3$ \rtwo{(which corresponds to simplifying\ \eqref{eq:uOmEx} to separate the angle-dependent motions from pure translational ones, see\ \eqref{eq:uOmSimp})}. \rone{For translational velocity, the finest feasible discretization is not sufficiently refined to see complete saturation of $\norm{\Usb-\V{U}}$ (see Fig.\ \ref{fig:TransRotk}), so} we report error bars in Fig.\ \ref{fig:TransRotErk} that give the difference in $\norm{\Usb-\V{U}}$ between our last two levels of refinement (\rone{the last two data points on each curve in Fig.\ \ref{fig:TransRotk}}). Based on Fig.\ \ref{fig:TransRotk}, the numerical error in the BI calculations likely causes an underestimate of the SBT error for $k=2$ and $k=2.5$ \rone{(since the error increases from the second-finest to finest discretization)}, while for $k=2.85$ and $k=3$ we likely overestimate the SBT error  \rone{(since the error decreases from the second-finest to finest discretization)}. 

\rtwo{To find the approximately optimal value for $k$, we systematically vary $k$ between 2 and 3 in increments of 0.05, with the step size being relatively large because we are constrained by the accuracy of our boundary integral calculations. In Fig.\ \ref{fig:TransRotErk}, we show the SBT errors for $k=2, 2.5$, and 3, which represent evenly spaced values between 2 and 3, and compare them to our optimal value of $k=2.85$, which has the smallest errors for $\epsc \leq 0.01$. } For $k=2$ (which is our first guess from the first line of the rotlet integral expansion\ \eqref{eq:uOmEx}), and $k=2.5$, we observe an error which saturates to a constant as we decrease $\epsc$, meaning that the error in the SBT\ \eqref{eq:SBTtrans} when $k=2$ or $2.5$ is asymptotically $\mathcal{O}(1)$. Increasing to $k=2.85$, we observe saturated errors that are roughly proportional to $\epsc$, which is exactly what we want to achieve in SBT. When we increase again to $k=3$, we see errors which initially decrease with $\epsc$, but appear to plateau to an $\mathcal{O}(1)$ value. It appears then, that there is an optimal $k$ around $k=2.85$ which makes the SBT formula\ \eqref{eq:SBTtrans} accurate to $\mathcal{O}(\epsc)$. 

\rone{To verify that $k=2.85$ is independent of the fiber geometry, in Fig.\ \ref{fig:TransRotErk} we show as a light blue line the results when we perform the exact same test on a fiber with tangent vector
\begin{equation}
\label{eq:Geo2}
\Xs(s) = \frac{1}{\sqrt{2}}\left(\cos{\left(qs^3(s-L)^3\right)}, \sin{\left(qs^3 (s-L)^3\right)}, 1\right),
\end{equation}
with the position obtained by integration. This fiber forms an ``S'' shape, which is different from the half turn helix of\ \eqref{eq:Xhelix} (see the left panel of Fig.\ \ref{fig:TransRotErk}). While the shape is different, Fig.\ \ref{fig:TransRotErk} shows that using $k=2.85$ is once again sufficient to match the SBT theory\ \eqref{eq:SBTtrans} with the three-dimensional boundary integral calculation, which suggests that the optimal $k$ is independent of the fiber shape.}

While this is by no means a proof that the correct rot-trans coupling coefficient\footnote{Indeed, there is no \emph{a priori} reason to believe\ \eqref{eq:SBTtrans} or a similarly simple equation holds rigorously from\ \eqref{eq:Uwant} and\ \eqref{eq:UBI}--\eqref{eq:nparTrac}.} in\ \eqref{eq:SBTtrans} is $k=2.85$, it does provide us with a good educated guess that can be used to cheaply approximate boundary integral simulations with small error, and get an idea of how close our RPY formulas are to the true three-dimensional dynamics. In particular, using an RPY radius of $\eps \approx 1.86 \rc$ in\ \eqref{eq:RotTransRPY} gives the formula\ \eqref{eq:kRT2} with $k=2.85$. This value of the RPY radius is comparable, but fairly larger than, what we obtained by matching the trans-trans ($\eps_\text{tt} \approx 1.12 \rc$) and rot-rot ($\eps_\text{rr} \approx 1.06 \rc$) mobilities in Section\ \ref{sec:compareRPSB}.

\section{Conclusion}
We examined the deficiencies of existing slender body theories (SBTs) for rotating filaments, and then proposed a semi-empirical SBT for tubular filaments derived from line integrals of Rotne-Prager-Yamakawa (RPY) singularities. We showed that the singularity-based SBTs of Keller+Rubinow and Johnson \cite{krub, johnson}, \rone{which were designed with the resistance problem in mind,} both suffer from the same issues for \rone{rotation-translation coupling in the mobility problem}: when the fiber is driven by an applied torque at its base, their theories imply that torque induces an $\mathcal{O}(1)$ angular-dependent translational velocity on the fiber cross section, which violates the SBT boundary condition that the fiber only translates and rotates. In addition, these previous SBTs exhibit a lack of symmetry in the $2 \times 2$ mobility operator because of assumptions about the relative order of magnitude of the fiber translational and rotational velocities. As we showed previously \cite[Sec.~5.3]{maxian2022hydrodynamics}, neglecting rotation-translation coupling can lead to overestimation of the torque required to induce large-scale filament deformations. It is therefore important to derive an SBT which properly treats rotation-translation coupling.

As a workaround to the true problem of obtaining an SBT for a three-dimensional cylindrical (or prolate-spherioidal) filament, we used a different definition of a filament as a line of infinitely many \rthr{Rotne-Prager-Yamakawa (RPY)} regularized singularities, similar to what has been done previously by Cortez and Nicholas \rtwo{for regularized Stokeslets} \cite{cortez2012slender}. We then defined each component of the $2 \times 2$ mobility operator (translation-translation, rotation-translation, etc.)\ as a line integral of the corresponding fundamental regularized singularity given in\ \eqref{eq:Ukernels}. Because these regularized singularities, which define the mobility for a pair of ``blobs,'' give a symmetric $2 \times 2$ mobility, our mobility for fibers is automatically symmetric, which alleviates one of the two major problems with previous SBTs. The other issue, the $\mathcal{O}(1)$ angular dependence of the translational velocity on a cross section, is sidestepped because we use an isotropic regularized singularity function (a delta function on a sphere). This comes at a price: by using a series of spheres to represent the fiber centerline, we lose fidelity to the true three-dimensional geometry \rone{and the fluid flows near the fiilament surface}. While this is far from ideal, it gives a mobility which is physical and simple to compute, since all that results is a local drag term plus a finite part integral for the trans-trans, rot-trans, and trans-rot terms, for which special quadrature schemes have already been developed \cite{tornquad}. 

Using matched asymptotics on the RPY kernels, we found a $2 \times 2$ mobility relationship, which we then compared to SBT for a three-dimensional tubular fiber. For translation-translation and rotation-rotation, we found that the two mobilities take the same form, but with an $\mathcal{O}(1)$ difference in the constants. The discrepancy in the constants can be removed via a judicious choice of regularized singularity radius, which is different for each component of the mobility; $\eps \approx 1.06 \rc$ gives a perfect match in the rot-rot mobilities, while $\eps \approx 1.12 \rc$ matches the trans-trans mobilities. We used the correspondence between RPY asymptotics and SBT to posit the relationship\ \eqref{eq:kRT2} for the off-diagonal rotation-translation components in SBT. The relationship has a single unknown coefficient $k\approx 2.85$, which we obtained in Section\ \ref{sec:BI} from boundary integral calculations. The value of $k$ we found corresponds to a regularized singularity radius $\eps \approx 1.86 \rc$. 

Using the RPY kernels simplified a number of potential issues that \rone{arise for more rigorous models of the three-dimensional fiber geometry}. First, we were able to easily derive mobility formulas near the fiber endpoints simply by modifying the domain of integration in the inner expansion (see Appendix\ \ref{sec:EPs}). \rone{This is in contrast to the more rigorous three-dimensional case, where a nonsingular endpoint flow is obtained by assuming an ellipsoidally-decaying radius function and distributing the centerline singularities only between the generalized foci of the body \cite{johnson}.} Furthermore, the RPY mobility immediately generalizes to multiple filaments, since the flow induced on filament $j$ can also be computed by an integral of the RPY kernel over filament $i$. In SBT, while the pointwise velocity induced by singularities might be constant on a cross section of fiber $i$, the same might not be true on a nearby fiber $j$, which leads to an ambiguity in how the induced flow on the centerline of fiber $j$ is assigned. \rone{This is especially relevant when fibers are in near contact, and shows why only three-dimensional boundary integral approaches are precise when fibers are $\mathcal{O}(\epsc)$ apart; such calculations are however very complex and expensive for suspensions of many fibers \cite{yan2022toward}.} 

Our boundary integral calculations in Section\ \ref{sec:BI} can be viewed as both a check on the accuracy of the rotation-translation coupling terms in our RPY theory, and also as a semi-empirical SBT for translation-rotation coupling. Indeed, these calculations are an attempt to find an SBT by reversing the process Koens and Lauga \cite{koens2018boundary} used to show that translational SBT can be derived from the BI equation. While we showed that this approach is unworkable analytically for the rotation-translation coupling problem, numerical calculation allowed us to extract the relationship
\begin{equation}
\label{eq:SBTEmp}
8 \pi \mu \tr{\V{U}} \approx \left(\ln{\left(\epsc^{-2}\right)}-2.85\right) \left(\Xs \times \ds \Xs\right)\frac{n^\parallel}{2}+\tr{\V{U}}^{\text{(FP)}}+\mathcal{O}(\epsc)
\end{equation}
for the translational velocity due to parallel torque on an ellipsoidal filament, with the finite part integral $\tr{\V{U}}^{\text{(FP)}}$ defined in\ \eqref{eq:UnFP}. Of course, this result is an estimate for a particular filament, and there is no way to show it is universal, although tests \rone{on multiple geometries gave the same $\mathcal{O}(1)$ constant of $2.85$}. Still, our guess for the rotation-translation coupling tensor is informed by the asymptotics of the RPY tensor, and so, if it is incorrect or non-universal, the RPY mobility\ \eqref{eq:RotTransRPY} must miss some of the essential physics in the modeling of slender filaments. Given the number of studies that have used the RPY tensor to model filament hydrodynamics (e.g., \cite{wang2012flipping, keavRPY, kallemov2016immersed, westwood2021coordinated}), this would be concerning. 

Our work here is by no means the final say on SBT for twisting filament, but is rather intended as a starting point to provoke future research in three main directions. First, there is the possibility of developing a singularity representation for the flow around a twisting filament that properly matches the velocity on the fiber cross section. For an ellipsoidal filament, the velocity on a cross section due to a rotlet\ \eqref{eq:uOmEx} can be expanded using\ \eqref{eq:ExpandAng} as 
\begin{gather}
\label{eq:uOmSimp}
8\pi \mu \V{u}^{(\Psi)}\left(\Xhat\left(s,\theta\right)\right) = 
\frac{2 n^\parallel}{\rc}\left(\Xs \times \hat{\V{r}}\right) +\frac{n^\parallel}{2} \left(\ln{\left(\epsc^{-2}\right)}-3\right) \left(\Xs \times \ds \Xs\right)+\tr{\V{U}}^{\text{(FP)}}\\ \nonumber 
 +\frac{n^\parallel \kappa}{2}\left(\eb \cos{2\theta} -\en \sin{2\theta}\right)+
\mathcal{O}\left(\epsilon\right).
\end{gather}
Thus, the task is to find another singularity (or set of singularities) that can eliminate the $\mathcal{O}(1)$ dependence on $2\theta$ in the second line, thereby making the cross-sectional velocity a pure rotation (the $\Xs \times \rhat$ term) plus a pure translation (the next two terms in the first line). Based on our empirical SBT\ \eqref{eq:SBTEmp}, we expect such a singularity, if it exists, to eliminate the angular dependence while introducing a small correction to the $\mathcal{O}(1)$ term multiplying $\Xs \times \ds{\Xs}$. 

Given that it is difficult to find such a singularity representation, if it exists at all, our empirical approach in Section\ \ref{sec:BI} offers some promise, if only we could reach even smaller $\epsc$. In the case when $\Psi^\parallel \sim 1/\epsc^2$, a more accurate numerical approach is necessary to do so, since in that case the surface tractions scale like $1/\epsc$, and it becomes difficult to confirm $\mathcal{O}(\epsc)$ scaling of asymptotic expressions. And finally, there is an opportunity to analyze what Stokes-related partial differential equation such a slender body theory is the solution of (this is an extension of the work of Mori, Ohm, and others for translation \cite{mori2018theoretical,mori2020accuracy, morifree}). 

\section*{Acknowledgments}
We thank Ajeet Kumar and Mohit Garg for useful discussions that led to this work. This work was supported by the National Science
Foundation through Research Training
Group in Modeling and Simulation under award RTG/DMS-1646339 and through the Division of Mathematical Sciences award DMS-2052515. 

\subsection*{Declaration of Interests}
The authors report no conflict of interest.

\begin{appendices}
\setcounter{equation}{0}
\renewcommand{\theequation}{\thesection.\arabic{equation}}

\setcounter{equation}{0}
\section{Isotropic single layer expansion \label{sec:KLAppen}}
To expand the isotropic single layer\ \eqref{eq:Ui1}, we begin with an inner expansion of the displacement $\V{R}(s,s',\theta,\theta')=\widehat{\V{X}}(s,\theta)-\widehat{\V{X}}(s',\theta')$,
\begin{gather}
\label{eq:Rexp}
\V{R}(s,s',\theta,\theta')=\widehat{\V{X}}(s,\theta)-\widehat{\V{X}}(s',\theta')= \V{X}(s)-\V{X}(s') + \er(s,\theta) - \er(s',\theta'),
\end{gather}
where $\er(s,\theta)$ is the unit vector going from $\V{X}(s)$ to $\widehat{\V{X}}(s,\theta)$. Since the geometry is more complicated in this true three-dimensional case, it is helpful to parameterize $\er$ by an angle $\theta_i(s)$ satisfying $\ds \theta_i=\eb \cdot \ds\en$ (the curve torsion) \cite[Eq.~(2.2)]{koens2018boundary}. This gives \cite[Sec.~2]{koens2018boundary}
\begin{gather}
\label{eq:er}
\er(s,\theta) = \en \cos{(\theta-\theta_i(s))} + \eb \sin{(\theta-\theta_i(s))} \\ \nonumber
\ds{\er}(s,\theta) = -\kappa(s) \cos{(\theta-\theta_i)}\Xs(s)
\end{gather}
Substituting into\ \eqref{eq:Rexp}, we can now write the first two terms in the Taylor expansion of $\V{R}$ and $1/R$ as 
\begin{gather}
\nonumber
\V{R}\left(\widehat{\V{X}}(s,\theta),\widehat{\V{X}}(s',\theta')\right) \approx \rc  \left(-\xi \Xs + \Delta \er(s,\theta,\theta')\right)+\rc^2 \left(\frac{\xi^2}{2} \ds\Xs-\xi \kappa  \cos{(\theta'-\theta_i)}\Xs\right) \\ \label{eq:Rfinal}
\frac{1}{R\left(\widehat{\V{X}}(s,\theta),\widehat{\V{X}}(s',\theta')\right)} = \frac{1}{\rc \sqrt{\xi^2+\Delta \er \cdot \Delta \er}}+\frac{-\xi^2 \kappa \cos{(\theta'-\theta_i)} + \xi^2/2 \Xs' \cdot \Delta \er}{2\sqrt{\xi^2+\Delta \er\cdot  \Delta \er}^3}
\end{gather}
where $\Delta \er(s,\theta,\theta')=\er(s,\theta)-\er(s,\theta')$ and $\xi = (s'-s)/\rc$ is order 1. Note the appearance of the centerline curvature (scalar $\kappa$ and vector $\ds{\Xs}$) at this order, which is vital to yield proper rot-trans coupling. Combining\ \eqref{eq:Rfinal} with the Taylor series representation $\V{f}(s',\theta')=\V{f}(s,\theta')+\rc \xi \ds \V{f}(s,\theta')$, we have the inner expansion of the integrand in\ \eqref{eq:Ui1B4} as
\begin{gather}
\label{eq:fst}
\frac{\V{f}(s',\theta')}{R} = \frac{\V{f}(s,\theta')}{\rc \sqrt{\xi^2+\Delta \er \Delta \er}}+ \frac{\xi \ds{\V{f}}(s,\theta')}{\sqrt{\xi^2+\Delta \er \Delta \er}}\\ \nonumber+\frac{-\xi^2 \kappa  \cos{(\theta'-\theta_i)} + \xi^2/2 (\ds \Xs \cdot \Delta \er)}{2\sqrt{\xi^2+\Delta \er \Delta \er}^3}\V{f}(s,\theta')+\mathcal{O}(\epsc).
\end{gather}
The second line already makes it clear how rot-trans coupling is going to arise. If $\V{f}(s,\theta')=\V{f}(s)$ (no angular dependence), then there will still be a term in the velocity proportional to $\cos{\theta}\V{f}(s)$, which is rotational velocity coming from translational force, or rot-trans coupling. 

To complete the expansion, we proceed with the integration of\ \eqref{eq:fst} from $\xi=-s/\rc$ to $\xi=(L-s)/\rc$. The integration can be done exactly, and then expanded in $\epsc$ to yield simpler expressions \cite[Table~1]{koens2018boundary}. After using\ \eqref{eq:er} to determine $\ds \Xs \cdot \Delta \er=0$, then integrating what remains, we have  
\begin{gather}
\label{eq:Uinttheta}
\V U_{i1}(s,\theta)=\int_{0}^{2\pi} \Bigg{(} \ln{\left(\frac{4s(L-s)}{\rc^2 (\Delta \er \cdot \Delta \er)}\right)}\V{f}(s,\theta') + \left(L-2s\right) \ds\V{f}(s,\theta)\\ \nonumber - \frac{\epsc \kappa \cos{\left(\theta'-\theta_i(s)\right)}\V{f}(s,\theta')}{2} \left( \ln{\left(\frac{4s(L-s)}{\rc^2 (\Delta \er \cdot \Delta \er)}\right)}-2\right)\Bigg{)} \, d\theta'+\mathcal{O}(\epsc^2).
\end{gather}
We substitute the one-term Fourier series representation\ \eqref{eq:OneTrac} into\ \eqref{eq:Uinttheta} and integrate over $\theta'$ to obtain\ \eqref{eq:Ui1}. To integrate, we use $\Delta \er \cdot \Delta \er =2\left(1-\cos{\left(\theta-\theta'\right)}\right)$.

\setcounter{equation}{0}
\section{Inner expansions for RPY asymptotics \label{sec:EPs}}
This appendix is devoted to the inner expansions of the RPY velocities\ \eqref{eq:Uconv}. Using the rescaled variable $\xi$ defined in\ \eqref{eq:xidef}, we write the series expansions
\begin{gather}
\label{eq:innerAsymp}
\V{R}(s') = \V{X}(s)-\V{X}(s') = -\xi \eps \Xs(s)-\frac{\xi^2 \eps^2}{2}\ds \Xs(s)+\mathcal{O}\left(\eps^3\right)\\ \nonumber 
\V{R}\V{R} = \xi^2 \eps^2 \Xs(s) \Xs(s) +\mathcal{O}\left(\eps^3\right),\\[2 pt]   \nonumber
R^2 = \V{R} \cdot \V{R} = \xi^2 \eps^2 + \mathcal{O}\left(\eps^3\right), \quad R = |\xi|\eps+\mathcal{O}\left(\eps^2\right),\\[2 pt] \nonumber
R^{-1} = \frac{1}{|\xi|\eps}+\mathcal{O}(1), \quad 
R^{-3} = \frac{1}{|\xi|^3 \eps^3} + \mathcal{O}\left(\eps^{-2}\right), \quad 
R^{-5} = \frac{1}{|\xi|^5 \eps^5} + \mathcal{O}\left(\eps^{-4}\right). \\[2 pt]  \nonumber
\V{f}\left(s'\right) = \V{f}(s) + \xi \eps \ds{\V{f}}(s)+\mathcal{O}(\eps^2) \qquad  n^\parallel \left(s'\right) = n^\parallel (s) + \xi \eps \ds{n^\parallel }(s)+\mathcal{O}(\eps^2),
\end{gather}
which we substitute into the corresponding RPY velocity formulas\ \eqref{eq:rpyint},\ \eqref{eq:tfromt},\ \eqref{eq:rfromf}, and\ \eqref{eq:rfromt}. What results is two separate integrals over $\xi$: one for $|\xi| > 2$ (when the RPY spheres do not overlap), and another for $|\xi| \leq 2$ (when they do). The bounds on the integrals are
\begin{gather}
\label{eq:intBounds}
\int_{|\xi| > 2} f(\xi) \, d\xi = 
\begin{cases}
\int_{-s/\eps}^{-2} f(\xi) \, d\xi +\int_2^{(L-s)/\eps} f(\xi) \, d\xi & 2\eps < s < L-2\eps\\
\int_2^{(L-s)/\eps} f(\xi) \, d\xi & s \leq 2\eps\\
\int_{-s/\eps}^{-2} f(\xi) \, d\xi & s \geq L-2\eps\\
\end{cases} \\ \nonumber
\int_{|\xi| \leq 2} f(\xi) \, d\xi = 
\begin{cases}
\int_{-2}^{-2} f(\xi) \, d\xi & 2\eps < s < L-2\eps\\
\int_{-s/\eps}^2 f(\xi) \, d\xi & s \leq 2\eps\\
\int_{-2}^{(L-s)/\eps} f(\xi) \, d\xi & s \geq L-2\eps\\
\end{cases}
\end{gather}
With these integration bounds, the inner velocity can be computed for any $s$ on the filament, as we do next.

\subsection{Translation from force}
Beginning with the trans-trans mobility\ \eqref{eq:rpyint}, we start with the part of the integral in the region $R> 2\eps$. For this we first need to integrate the inner expansion of the Stokeslet
\begin{align}
\label{eq:sletin}
\int_{R > 2\eps} \Slet{\V{X}(s),\V{X}(s')}&\V{f}\left(s'\right) \, ds' \approx \int_{|\xi| > 2}  \left(\frac{\M{I}+\Xs(s) \Xs(s)}{|\xi| \eps}\right)\V{f}(s)  \eps \, d\xi \\[2 pt] \nonumber
& = \left(\M{I}+\Xs(s)\Xs(s)\right)\V{f}(s)\int_{|\xi| > 2} \frac{1}{|\xi|} d\xi.
\end{align}
Unlike the Stokeslet, the doublet term is only included in the inner expansion. Its expansion is given by
\begin{align}
\label{eq:dbletin}
\frac{2\eps^2}{3} & \int_{R > 2\eps} \Dlet{\V{X}(s),\V{X}(s')} \, ds' \approx \frac{2 \eps^2}{3} \int_{|\xi| > 2}  \left(\frac{\M{I}-3\Xs(s) \Xs(s)}{|\xi|^3 \eps^3}\right)\V{f}(s)  \eps \, d\xi \\[4 pt]
\nonumber
& = \frac{2\left(\M{I}-3\Xs(s) \Xs(s)\right)\V{f}(s)}{3}  \int_{|\xi| > 2} \frac{1}{|\xi|^3} \, d\xi .
\end{align}
It still remains to include in the inner expansion the term for $R \leq 2\eps$. For this we have the inner expansion of the first term
\begin{align}
\label{eq:rl1}
\int_{R \leq 2\eps} \left(\frac{4}{3\eps}-\frac{3R\left(s'\right)}{8\eps^2}\right)\V{f}\left(s'\right) \, ds'  &\approx\V{f}(s)\int_{|\xi| \leq 2}  \left(\frac{4}{3}-\frac{3|\xi|}{8}\right) \, d\xi .
\end{align}
Likewise the second term for $R \leq 2\eps$ has the inner expansion
\begin{align}
\label{eq:rl2}
 \int_{R \leq 2\eps} \frac{1}{8\eps^2R\left(s'\right)} \left(\V{R}\V{R}\right)\left(s'\right)\V{f}\left(s'\right) \, ds' & \approx 
\Xs(s)\Xs(s)\V{f}\left(s\right)\int_{|\xi| \leq 2} \frac{ |\xi|}{8} \, d\xi.
\end{align}
Adding the terms\ \eqref{eq:sletin},\ \eqref{eq:dbletin},\ \eqref{eq:rl1}, and\ \eqref{eq:rl2}, and computing integrals using the bounds\ \eqref{eq:intBounds}, we obtain the complete inner expansion
\begin{gather}
\label{eq:UinnerA}
8 \pi \mu \tt{\V{U}}^{(\text{inner})}(s) =  
\left(a_L(s) \left(\M{I}+\Xs(s)\Xs(s)\right)
+a_I(s)\M{I} +a_\tau(s) \Xs(s)\Xs(s)\right)\V{f}(s)\\ \nonumber
a_L(s) = \begin{cases}
\ln{\left(\dfrac{(L-s)s}{4\eps^2}\right)} & 2\eps < s < L-2\eps\\[4 pt]
\ln{\left(\dfrac{(L-s)}{2\eps}\right)}& s \leq 2\eps \\[4 pt]
\ln{\left(\dfrac{s}{2\eps}\right)} & s \geq L-2\eps
\end{cases} \\[4 pt] \nonumber
a_I(s) = \begin{cases}
4-\dfrac{\eps^2}{3s^2}-\dfrac{\eps^2}{3(L-s)^2}& 2\eps < s < L-2\eps\\[4 pt]
2+\dfrac{4s}{3\eps} - \dfrac{3s^2}{16\eps^2}-\dfrac{\eps^2}{3(L-s)^2}& s \leq 2\eps \\[4 pt]
2+\dfrac{4(L-s)}{3\eps} - \dfrac{3(L-s)^2}{16\eps^2}-\dfrac{\eps^2}{3s^2}& s \geq L-2\eps
\end{cases}
a_\tau(s) = \begin{cases}
\dfrac{\eps^2}{s^2}+\dfrac{\eps^2}{(L-s)^2} & 2\eps < s < L-2\eps\\[4 pt]
\dfrac{s^2}{16\eps^2}+\dfrac{\eps^2}{(L-s)^2} & s \leq 2\eps \\[4 pt]
\dfrac{(L-s)^2}{16\eps^2}+\dfrac{\eps^2}{s^2}& s \geq L-2\eps
\end{cases}
\end{gather}
The velocity $\tt{\V{U}}^{(\text{inner})}$ is twice continuously differentiable, since it is the integral of a function which is once continuously differentiable. 


\subsection{Rot-trans and trans-rot coupling\label{sec:rpyttorq}}
We now move on to the inner expansion in the case of translation-rotation coupling\ \eqref{eq:tfromt}. Beginning with the rotlet term for $R > 2\eps$, we write the rotlet in terms of the inner variables as
\begin{align}
\label{eq:ncrossR}
\Xs(s') \times \V{R}(s') & \approx \frac{\xi^2 \eps^2}{2}\left(\Xs(s) \times \ds{\Xs}(s)\right)\\
\label{eq:inner2b}
\int_{R > 2\eps} \frac{\Xs(s') \times \V{R}(s')}{R(s')^3}n^\parallel(s') \, ds' &\approx 
\int_{|\xi| > 2}  \frac{\left(\Xs(s) \times \ds{\Xs}(s)\right)}{2|\xi| \eps} n^\parallel(s) \eps \, d\xi .
\end{align}
We next need to account for the term for $R \leq 2\eps$, or the second term in\ \eqref{eq:tfromt}. Using\ \eqref{eq:ncrossR}, we have that
\begin{align}
\label{eq:intclosett}
\frac{1}{2\eps^2} \int_{R < 2\eps} \left(\frac{1}{\eps}-\frac{3R}{ 8\eps^2}\right)\left(\V{n}(s') \times \V{R}(s')\right) \, ds'&\approx
\frac{\left(\Xs(s) \times \ds{\Xs}(s)\right)}{4 \eps}\int_{|\xi| < 2} \left(1-\frac{3}{8}|\xi| \right)\xi^2  \eps \, d\xi.
\end{align}
Adding\ \eqref{eq:inner2b} and\ \eqref{eq:intclosett} and using\ \eqref{eq:intBounds} to compute the inner expansion integrals, we obtain
\begin{gather}
\label{eq:UinTTA}
8\pi \mu \tr{\V{U}}^\text{(inner)}(s) = \frac{n^\parallel(s)}{2}\left(\Xs(s)\times \ds \Xs (s)\right)  c_c(s) \\[4 pt] \nonumber
c_c(s) = \begin{cases}
\ln{\left(\dfrac{(L-s)s}{4\eps^2}\right)}+\dfrac{7}{6} & 2\eps < s < L-2\eps\\[4 pt]
\ln{\left(\dfrac{(L-s)}{2\eps}\right)}+\dfrac{7}{12}+\dfrac{s^3}{6\eps^3}-\dfrac{3s^4}{64\eps^4}& s \leq 2\eps \\[4 pt]
\ln{\left(\dfrac{s}{2\eps}\right)}+\dfrac{7}{12}+\dfrac{(L-s)^3}{6\eps^3}-\dfrac{3(L-s)^4}{64\eps^4}& s \geq L-2\eps
\end{cases}
\end{gather}
which is a twice continuously differentiable function. 

The inner expansion of angular velocity from force is given symmetrically as
\begin{gather}
\label{eq:matchRFApp}
8\pi \mu \rt{\Psi}^{\parallel,\text{(inner)}}(s) = \frac{1}{2}c_c(s)\left(\Xs(s) \times \ds{\Xs}(s)\right) \cdot \V{f}(s).
\end{gather}

\subsection{Rotation from torque}
The inner expansion for\ \eqref{eq:rfromt} contains two parts: the doublet, and terms for $R \leq 2\eps$. We have already seen the expansion of the doublet in\ \eqref{eq:dbletin}. Multiplying the result in\ \eqref{eq:dbletin} by the appropriate prefactors, we obtain
\begin{gather}
\label{eq:term1}
-\frac{1}{2}\int_{R > 2\eps } \left(\frac{\V{n}(s')}{R(s')^3}-3\frac{\V{R}(s')\left(\V{R}(s') \cdot \V{n}(s')\right)}{R(s')^5}\right) \, ds' \approx  -\frac{\left(\M{I}-3\Xs(s) \Xs(s)\right)\V{f}(s)}{2\eps^2}  \int_{|\xi| > 2} \frac{1}{|\xi|^3} \, d\xi,
\end{gather}
which which is accurate to $\mathcal{O}\left(\epsRS^{-1}\right)$,  although in the fiber interior that term actually cancels, leaving an $\mathcal{O}\left(\log{\epsRS^{-1}}\right)$ dependence.

In a similar way, the inner expansions of the $R \leq 2\eps$ terms in the fiber interior are
\begin{gather}
\label{eq:term2}
\frac{1}{\eps^3}\int_{R < 2\eps} \left(1-\frac{27R(s')}{32\eps}+\frac{5 R(s')^3}{64 \eps^3}\right)\V{n}(s') \, ds'  \approx  \frac{\V{n}(s)}{\eps^2}\int_{|\xi| \leq 2} \left(1-\frac{27}{32}|\xi|+\frac{5}{64}|\xi|^3 \right)\, d\xi \\
\nonumber
\frac{1}{\eps^3}\int_{R < 2\eps} \left(\left(\frac{9}{32aR(s')}-\frac{3R(s')}{64\eps^3}\right)\left(\V{R}\V{R}\right)(s')\right)\V{n}(s') \, ds'  \approx \frac{\Xs(s) \left(\Xs(s) \cdot \V{n}(s)\right)}{\eps^2} \int_{|\xi| \leq 2} \left(\frac{18-3|\xi|^2}{64}\right)|\xi| \, d\xi 
\end{gather}
with an error of $\mathcal{O}(\ln{\epsRS})$ in the fiber interior and $\mathcal{O}(\epsRS^{-1})$ at the endpoints.

Using the integral bounds\ \eqref{eq:intBounds} to integrate and sum\ \eqref{eq:term1} and\ \eqref{eq:term2}, the complete inner expansion for $\rr{\V{\Psi}}$ is given by
\begin{gather}
\label{eq:EPRRA}
8\pi \mu\rr{\V{\Psi}}^{\text{(inner)}}(s)=\left(p_I (s)\M{I}+p_\tau (s)\M{\Xs}(s)\Xs(s)\right)\V{n}(s), \qquad \text{where} \\[4 pt] \nonumber
p_I (s)= \dfrac{1}{\eps^2} \begin{cases}\dfrac{9}{8}+\dfrac{\eps^2}{4}\left(\dfrac{1}{s^2}+\dfrac{1}{(L-s)^2}\right) & 2 \eps < s < L-2\eps \\[6 pt]
\dfrac{9}{16}+\dfrac{s}{\eps}-\dfrac{27s^2}{64\eps^2}+\dfrac{5s^4}{256 \eps^4} +\dfrac{\eps^2}{4(L-s)^2}& s \leq 2 \eps \\[6 pt]
\dfrac{9}{16}+\dfrac{L-s}{\eps}-\dfrac{27(L-s)^2}{64\eps^2}+\dfrac{5(L-s)^4}{256 \eps^4} +\dfrac{\eps^2}{4s^2} & s \geq L-2\eps
\end{cases} \\[6 pt] \nonumber
p_\tau(s) = \dfrac{1}{\eps^2} \begin{cases}\dfrac{9}{8}-\dfrac{3\eps^2}{4}\left(\dfrac{1}{s^2}+\dfrac{1}{(L-s)^2}\right) & 2 \eps < s < L-2\eps \\[6 pt]
\dfrac{9}{16}+\dfrac{9s^2}{64\eps^2}-\dfrac{3s^4}{256 \eps^4} -\dfrac{3\eps^2}{4(L-s)^2}& s \leq 2 \eps \\[6 pt]
\dfrac{9}{16}+\dfrac{9(L-s)^2}{64\eps^2}-\dfrac{3(L-s)^4}{256 \eps^4} -\dfrac{3\eps^2}{4s^2}& s \geq L-2\eps
\end{cases} 
\end{gather}
which is twice continuously differentiable.

\section{Convergence of quadrature scheme \label{sec:BIConv}}
In this appendix, we discuss the convergence of the quadrature scheme for\ \eqref{eq:SingLayerS} for the rotating curved fiber in Section\ \ref{sec:FindK}. We report convergence as a function of $N_t=N N_\theta$ the total number of points on the filament boundary. In order to obtain smooth convergence, both $N$ and $N_\theta$ must be increased together. To maintain a similar spacing in the axial and circumferential directions, we use $N_\theta=16,20,\dots 32$, and $N \approx 0.25N_\theta/\epsc$.

\begin{figure}
\centering
\includegraphics[width=0.48\textwidth]{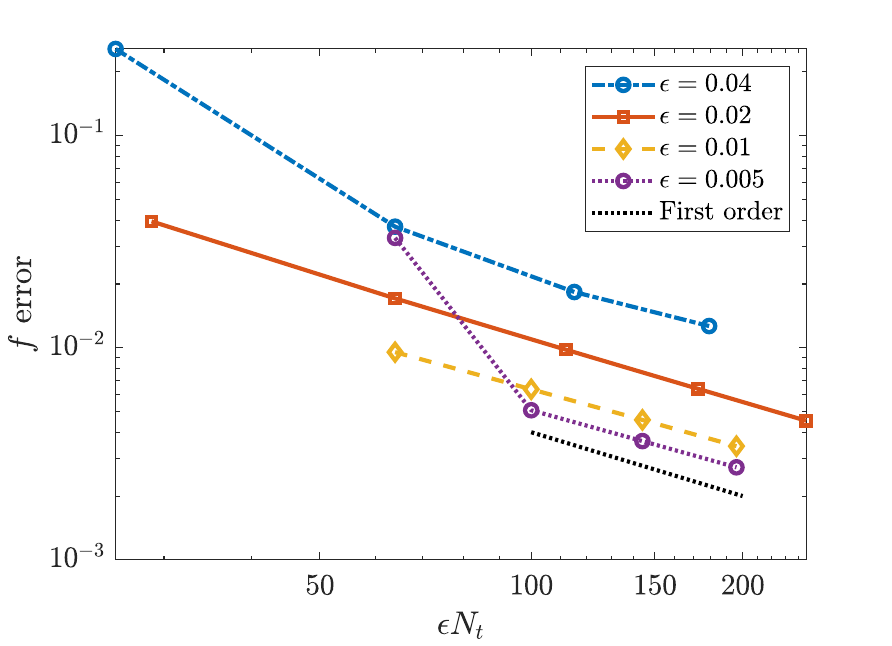}
\includegraphics[width=0.48\textwidth]{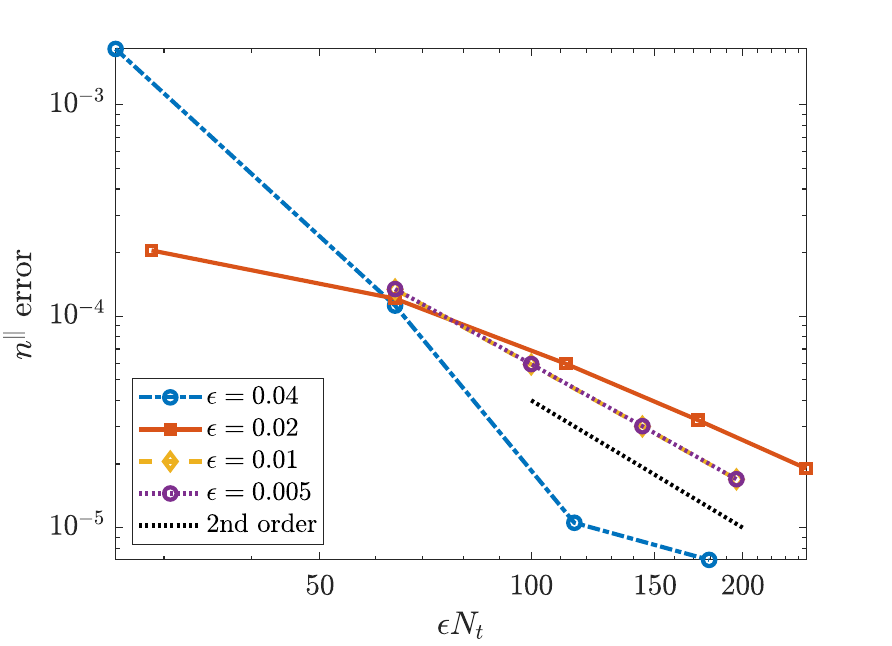}
\caption{\label{fig:TransRotConv} Convergence plot for the curved rotating filament\ \eqref{eq:Xhelix} studied in Section\ \ref{sec:FindK}. We show the self-convergence of our numerical method for (left) the force $\V{f}$ and (right) the parallel torque $n^\parallel$ on each cross section. The ``error'' is the $L^2$ difference of each quantity relative to the next level of refinement, normalized by the $L^2$ norm of the most refined solution. }
\end{figure}

We first perform a self-convergence study in Fig.\ \ref{fig:TransRotConv}, where we look at the $L^2$ errors in $\V{f}$ and $n^\parallel$, computed from\ \eqref{eq:fTrac} and\ \eqref{eq:nparTrac}, under refinement. For the force, the error at a given $N_t$ is defined as the $L^2$ difference of $\V{f}$ when compared to the next largest $N_t$, and similarly for $n^\parallel$. As $N_t$ increases, the force is asymptotically first-order accurate for every $\epsc$. The torque converges at a faster rate because, in this example, torque is dominated by local rotation-rotation dynamics. 

\begin{figure}
\centering
\includegraphics[width=0.6\textwidth]{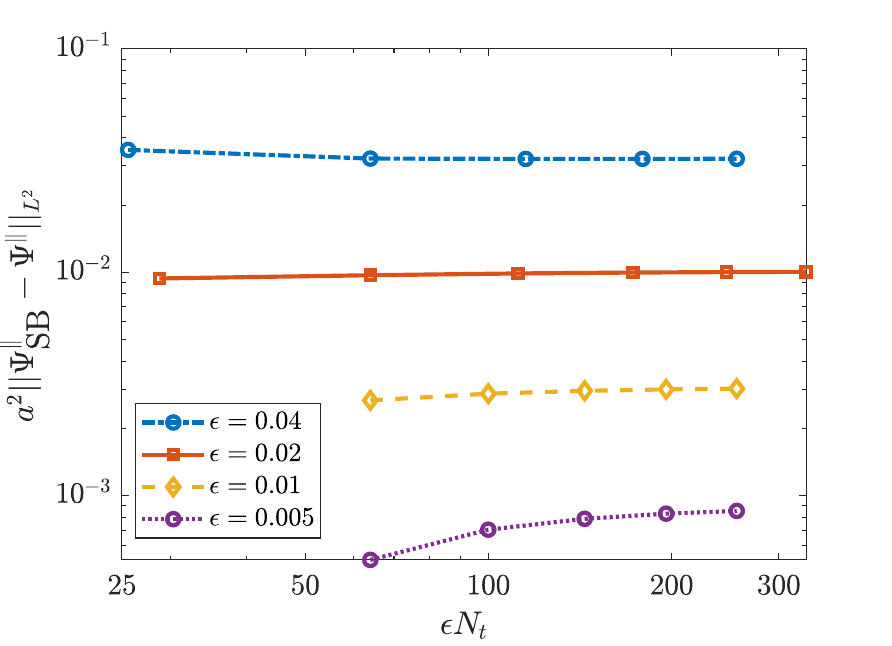}
\caption{\label{fig:RotTransRotk} SBT asymptotic errors for rotational velocity in the fully coupled problem. We show how the relative error $\rc^2\left(\Omsb-\Psi^\parallel\right)$ changes as we change $\epsc$ and the total number of discretization points $N_t$. The SBT rotational velocity $\Omsb$ is computed from\ \eqref{eq:kRT} and is dominated by rotation-rotation dynamics. It therefore does not depend on $k$, and so we show only a single $k=2.85$. }
\end{figure}

\begin{figure}
\centering
\includegraphics[width=\textwidth]{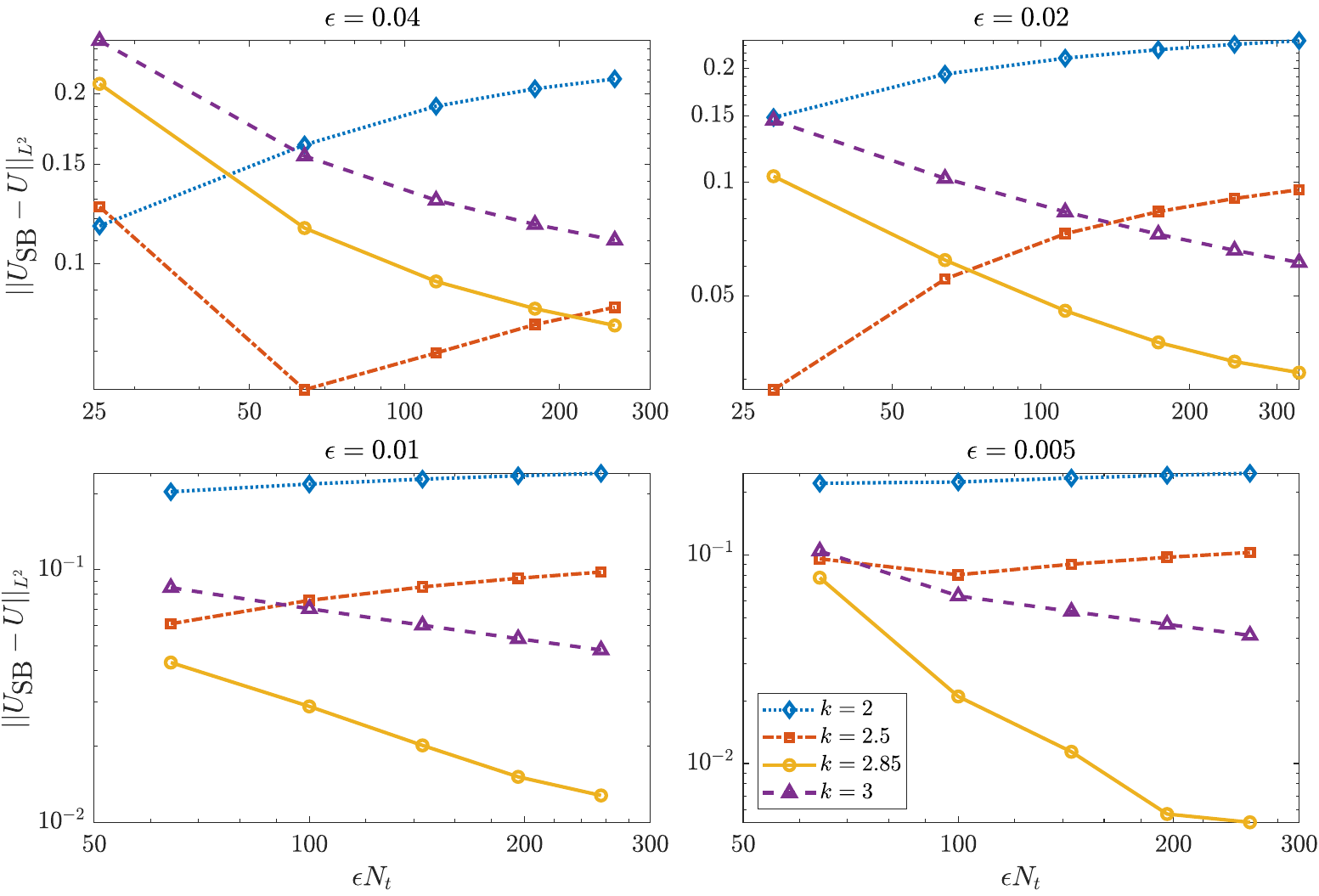}
\caption{\label{fig:TransRotk} SBT asymptotic errors for translational velocity in the fully coupled problem. We show how the error in the slender body velocity $\Usb$ changes as we refine the number of discretization points and the fiber aspect ratio. Here the velocity $\Usb$ is computed using\ \eqref{eq:SBTtrans} with the value of $k$ indicated in the legend. }
\end{figure}

 In Figs.\ \ref{fig:RotTransRotk} and\ \ref{fig:TransRotk}, we show how the convergence of the force and torque affects the convergence of the SBT errors, $\Usb-\V{U}$ and $\Omsb-\Psi^\parallel$ for various $\epsc$ and $k$. As we increase $N_t$, our hope is that the errors will saturate as we isolate the SBT asymptotic error from the boundary integral numerical error. For rotational velocity, which is dominated by torque and therefore independent of the $k$ used in\ \eqref{eq:kRT}, this is indeed the case, as we see constant errors under refinement in Fig.\ \ref{fig:RotTransRotk}. Furthermore, the relative error in $\Omsb$ scales roughly as $\epsc^2 \log{\epsc}$, which is expected since the rotational velocity is dominated by the rot-rot term. 

For translational velocity, we do not observe saturation of $\norm{\Usb-\V{U}}$ in Fig.\ \ref{fig:TransRotk}, but we can still obtain a good estimate of the SBT error by looking at our finest discretizations. In particular, the errors for $k=2$ and $k=2.5$ appear to be approaching a constant of about 0.2 and 0.1, respectively, irrespective of the particular value of $\epsc$. This indicates an $\mathcal{O}(1)$ error in the SBT asymptotic equation\ \eqref{eq:SBTtrans} for those values of $k$. However, when we increase $k$ further to 2.85 and later 3, the errors are much smaller, indicating that the SBT theory gets more accurate. To account for our BI solutions not being fully converged, we report error bars in Fig.\ \ref{fig:TransRotErk} that give the difference between our most refined and second-most refined errors in $\Usb$ (the difference between the rightmost and second-rightmost point on each colored curve in Fig.\ \ref{fig:TransRotk}).

\end{appendices}

\end{document}